\documentclass[%
 reprint,
 superscriptaddress,
 amsmath,amssymb,
 aps,
]{revtex4-2}

\usepackage{graphicx}% Include figure files
\usepackage{dcolumn}% Align table columns on decimal point
\usepackage{bm}% bold math
\usepackage{hyperref}% add hypertext capabilities
\usepackage{verbatim}
\usepackage[capitalise]{cleveref}
\usepackage[skins,theorems]{tcolorbox}
\usepackage{multirow}
\tcbset{highlight math style={enhanced,colframe=red,colback=white,arc=0pt,boxrule=1pt}}
\usepackage{kantlipsum}
\usepackage{xcolor}
\usepackage[normalem]{ulem}

\usepackage{lineno}
\usepackage{textcomp}
\def\qe{\textsc{Quantum ESPRESSO}\texttrademark}
\def\qeheat{\texttt{QEHeat}}
\def\sportran{\texttt{SporTran}}

\definecolor{tangerine}{rgb}{0.944,0.522,0}
\definecolor{verde}{rgb}{0.,0.6,0}
\definecolor{rosso}{rgb}{0.9,0.0,0.2}
\definecolor{arancio}{rgb}{0.9,0.6,0.4}
\definecolor{viola}{rgb}{0.9,0.3,0.9}
\definecolor{turchese}{rgb}{0,0.5,0.5}

\newcommand{\editor}[2]{%
  \expandafter\newcommand\csname #1note\endcsname[1]{%
    \textcolor{#2}{(\textbf{#1:} ##1)}}%
  \expandafter\newcommand\csname #1\endcsname[1]{%
    \textcolor{#2}{##1}}%
  \expandafter\newcommand\csname #1cancel\endcsname[1]{%
    \textcolor{#2}{\sout{##1}}}%
  \expandafter\newcommand\csname #1change\endcsname[2]{%
    \textcolor{#2}{\sout{##1} ##2}}%
  \newenvironment{#1text}{\color{#2}}{\color{black}}
}

\editor{C}{red}

\begin{document}

%\preprint{APS/123-QED}

\title{Heat transport in liquid water from first-principles and deep-neural-network simulations}%

\newcommand{\ai}{\textit{ab initio} }

\author{Davide Tisi}%
\affiliation{%
    SISSA -- Scuola Internazionale Superiore di Studi Avanzati, 34136 Trieste, Italy
 }%

\author{Linfeng Zhang}%
\affiliation{%
    Program in Applied and Computational Mathematics, Princeton University, Princeton, NJ 08544, USA
}%

\author{Riccardo Bertossa}%
\affiliation{%
    SISSA -- Scuola Internazionale Superiore di Studi Avanzati, 34136 Trieste, Italy
 }%
\author{Han Wang}%
\affiliation{%
    Laboratory of Computational Physics, Institute of Applied Physics and Computational Mathematics, Huayuan Road 6, Beijing 100088, People’s Republic of China
}%

\author{Roberto Car}%
\affiliation{%
    Program in Applied and Computational Mathematics, Princeton University, Princeton, NJ 08544, USA
}%
\affiliation{%
    Department of Chemistry, Department of Physics, and Princeton Institute for the Science and Technology of Materials, Princeton University, Princeton, NJ 08544, USA
}%

\author{Stefano Baroni}
\affiliation{%
    SISSA -- Scuola Internazionale Superiore di Studi Avanzati, 34136 Trieste, Italy
 }%
\affiliation{%
    CNR Istituto Officina dei Materiali, SISSA unit, 34136 Trieste, Italy
}%

\date{\today}% It is always \today, today,

\begin{abstract}
%\kant[1]
We compute the thermal conductivity of water within linear response theory from equilibrium molecular dynamics simulations,
by adopting two different approaches. In one, the potential energy surface (PES) is derived on the fly from the electronic ground state of 
density functional theory (DFT) and the corresponding analytical expression is used for the energy flux. 
In the other, the PES is represented by a deep neural network (DNN) trained on DFT data, whereby the PES
has an explicit local decomposition and the energy flux takes a particularly simple expression. 
By virtue of a gauge invariance principle, established by Marcolongo, Umari, and Baroni, 
the two approaches should be equivalent if the PES were reproduced accurately by the DNN model.     
We test this hypothesis by calculating the thermal conductivity, at the GGA (PBE) level of theory, using 
the direct formulation and its DNN proxy, finding that both approaches yield the same conductivity, in excess of the experimental 
value by approximately 60\%. Besides being numerically much more efficient than its direct DFT counterpart, the DNN 
scheme has the advantage of being easily applicable to more sophisticated DFT approximations, such as meta-GGA and hybrid
functionals, for which it would be hard to derive analytically the expression of the energy flux. We find in this way,
that a DNN model, trained on meta-GGA (SCAN) data, reduce the deviation from experiment of the predicted thermal conductivity by about
50\%, leaving the question open as to whether the residual error is due to deficiencies of the functional, 
to a neglect of nuclear quantum effects in the atomic dynamics, or, likely, to a combination of the two.

% molecular dynamics and from neural network potential. We benchmark the results against an \textit{ab initio} simulation at PBE gga level of theory, showing that the neural network can faithfully reproduce the DFT results at a computational cost comparable to classical dynamics. The PBE proved to overestimate the experimental results almost by $80$\%, thus we applied our scheme using a new neural network potential trained over accurate SCAN meta-GGA data \cite{Chen10846}. This last potential proved to reduce the bias but still couldn't get results compatible with experiment. Futures works will help us to understand the nature of this shift.
\end{abstract}

\maketitle

%\tableofcontents

\section{\label{sec:I}Introduction}
Heat transport plays an important role in many areas of science, such as, e.g., materials and planetary sciences, with major
impact on technological issues, such as energy saving and conversion, heat dissipation and shielding, etc.  
Numerical studies of heat transport at the molecular scale often rely on Boltzmann's kinetic approach \cite{Peierls1929,Klemens1958,Broido2007,Zhou2014}. This is adequate when the relaxation processes are dominated by binary collisions, 
as in the case of dilute gases of particles, such as atoms or molecules, or of quasiparticles, such as phonons in crystalline solids.  
A more general approach to calculate the transport coefficients is provided 
by simulations of the molecular dynamics (MD), either directly via non-equilibrium MD \cite{Evans2007,Allen2017,Muller-Plathe1997,Tenenbaum1982}, 
or in combination with Green-Kubo (GK) theory of linear response \cite{Green,Kubo,Evans2007,Allen2017} via equilibrium MD.
  
Much progress has been made in recent years to develop \emph{ab initio} approaches to heat transport based on electronic density functional 
theory (DFT). Some schemes used \emph{ad hoc} ingredients, such as a (rather arbitrary) quantum-mechanical definition of 
the atomic energies \cite{Kang2017}. Other schemes used a definition of the energy flux based on the normal-mode decomposition of the atomic coordinates and forces, which is only possible in crystalline solids \cite{Carbogno2017}. In this work we follow the formulation of Marcolongo, Umari, and Baroni (MUB) \cite{Marcolongo2016}, who derived a general DFT expression for the adiabatic energy flux, based on a  \emph{gauge invariance} principle for the transport coefficients \cite{Marcolongo2016,Ercole2016}. The MUB approach made \emph{ab initio} simulations of heat transport possible, not only for crystalline materials, but also for disordered systems, like liquids and glasses, albeit at the price of lengthy and costly simulations. Progress in statistical techniques for the analysis of the flux time series \cite{Ercole2017,Bertossa2019} made possible to 
achieve 10\% accuracy in the calculated thermal conductivity with simulations 
of a few dozen to a few hundred picoseconds. Still the computational burden of \emph{ab initio} MD, where the potential energy surface (PES) is generated on the fly from DFT, is heavy and requires access to high performance computer platforms 
for substantial wall-clock times (see, e.g., Appendix F of Ref. \cite{Marcolongo2021} for details on the computational cost of a MUB calculation).

In the last decade, a combination of standard electronic-structure methods, based on DFT, and new machine-learning techniques have allowed
the construction of inter-atomic potentials possessing quantum mechanical accuracy at a cost that is only marginally higher than that of classical force fields. All the machine learned potentials, which are represented either by 
a deep-neural network (DNN) \cite{Behler2007,Kondor2018,Smith2017,Linfeng2018} or by a Gaussian-process \cite{Bartok2010}, 
use a local decomposition of the total potential energy of the system in terms of atomic contributions, which makes 
straightforward to define the energy flux, or current, from which to compute the heat conductivity via GK theory.
%First principles studies of heat transport should benefit from such potential which construct representations of the many-body potential energy of the atoms, in terms of deep neural networks (DNN) \cite{Behler2007,Kondor2018,Smith2017,Linfeng2018} or of Gaussian processes \cite{Bartok2010} }

Here we adopt the recently developed deep potential (DP) framework \cite{NIPS2018_7696,Linfeng2018}. DP molecular dynamics (DPMD) simulations
have been used successfully to study bulk thermodynamic properties beyond the reach of direct DFT calculations \cite{DeepWater2021,Jiang2021,zhang2021modeling,Wu2021,Gartner26040,Niu2020,100_Milion_Atoms}, 
%\RCnote{In the above refs please correct hfO2 to HfO2, and jacob's to Jacob's. Also, I would include references reporting studies not  accessible at all with direct AIMD, such as T.E. Gartner et al., PNAS 117, 26040 (2020), and H. Niu et al., Nature Comm. 11, 1 (2020); also, rather than quoting ref. 28 D. Lu et al., it would be better to quote W. Jia et al. "Pushing the limit of molecular dynamics with ab initio accuracy to 100 million atoms with machine learning", SC20: International Conference for High Performance Computing, Networking, Storage and Analysis, pp. 1-14 (2020). Finally, when mentioning dynamic properties, DPMD has also been used successfully to compute infrared spectra of water and ice (ref. the paper on deep Wannier) as well as Raman spectra of water (ref paper by G.Sommers et al, PCCP (2020)}
as well as dynamic properties like mass diffusion in solid state electrolytes \cite{Marcolongo2019,huang2021deep}, thermal transport properties 
in silicon \cite{LIDeepMD}, infrared spectra of water and ice \cite{DeepWannier} and Raman spectra of water \cite{Sommers2020}.
In the present work, we report calculations of the thermal conductivity ($\kappa$) of water, a molecular liquid,
from both direct DFT and DPMD simulations. The close correspondence of the conductivities predicted with the two approaches
validates DPMD against the results obtained from the MUB current.
We adopt two popular DFT approximations: the PBE generalized gradient approximation (GGA) \cite{PBE} and the strongly constrained and appropriately normed (SCAN) meta-GGA \cite{SCANPerdew}. The SCAN functional describes water more accurately than PBE, relative to which it reduces the covalent character of the hydrogen bond and correctly predicts that the liquid is denser than the solid \cite{Chen10846}. However, 
expressions for the energy density and fluxes are not currently available for the SCAN functional, and its inherent complexity 
makes hard to derive usable analytical expressions for these quantities. Because of that, we used PBE to validate our methodology.
Our results show that direct DFT simulations based on the PBE functional, and simulations based on the corresponding DP model are in good agreement with each other, but distinctly overestimate the thermal conductivity relative to experiment. This outcome likely
reflects the well known tendency of PBE to overestimate the strength of the hydrogen bonds, enhancing short-range order and making 
liquid water more ``solid-like'' and prone to freezing \cite{Sit2005}. 
DPMD simulations trained on SCAN-DFT reduce substantially the error of the heat conductivity predicted by PBE, but do not eliminate it, 
thus leaving open the question as to its origin, which is possibly
due to residual deficiencies of the functional, to nuclear quantum effects
ignored in the MD equations of motion, or, likely, to a combination of the two.

The paper is organized as follows. In \cref{sec:Theory}, we recall the main aspects of the GK theory, along with two 
basic invariance principles of thermal transport that allow us, among other things, to define the MUB-DFT energy flux. 
In \cref{sec:deepMD}, we describe the DP model, derive the corresponding expression for the energy flux, and discuss 
the impact of the invariance principles within a DNN simulation framework. In \cref{sec:results}, we benchmark 
our DNN methodology against \emph{ab initio} MD simulations of liquid water at the PBE level of theory \cite{PBE}. 
Having proved that DPMD trustfully 
reproduces \emph{ab initio} results, in \cref{sec:SCAN}, we take advantage of 
the simple DNN expression for the heat current 
to compute the thermal transport coefficients of liquid water at the SCAN meta-GGA level of theory.
%\RCnote{The observation that SCAN would give a too complicated expression for the energy flux within DFT should be made earlier in the previous paragraphs, not here. Some additional explanation would be needed to avoid the impression that I get by reading this sentence that somewhere in the paper we do report the flux etc for SCAN, which turns out to be too complex. In fact, we do not even report it. I think that the logic should be 1. the PBE calculation is useful to establish how well DeePMD works in the MUB context. 2. having established that it works well and provides a substantially simpler expression for the energy flux etc, it can be used with confidence with other functionals like SCAN for which close expressions for the fluxes are not available}.
Section \ref{sec:Conclusions} contains our conclusions.

\section{\label{sec:Theory} Theory}
GK theory of linear response \cite{Green,Kubo} provides a rigorous and elegant framework to compute the atomic contribution to the thermal conductivity, $\kappa$, of extended systems, in terms of the stationary time series of the energy flux \cite{flux}, $\bm{J}^e$, evaluated at thermal equilibrium with MD. For an isotropic system of $N$ interacting particles, the GK expression for the heat conductivity reads:
\begin{equation}\label{GKeq}
    \kappa=\frac{V}{3k_BT^2}\int_0^{\infty} \langle \bm{J}^e(\Gamma_t) \cdot \bm{J}^e(\Gamma_0) \rangle dt,
\end{equation}
where $\Gamma_t$ indicates the time evolution of a point in phase space from the initial condition
% \RB{evolved for a time $t$ from its initial condition} 
$\Gamma_0$. 
% \RB{using the system's equation of motion}, $\langle \cdot \rangle$ represents an  \DTchange{equilibrium (canonical or microcanonical) average}{equilibrium average ( e.g. canonical or microcanonical, ...)} for the initial condition $\Gamma_0$, $k_B$ the Boltzmann constant, and $V$ and $T$ are the system's volume and temperature.
The definition of the energy current in \cref{GKeq} is the key ingredient for the computation of $\kappa$. This definition relies in general on extensivity, which allows the total, conserved, energy of an isolated system to be broken up into local contributions. In a classical setting, this is conveniently achieved by expressing the total energy as a sum of atomic energies, $\epsilon_n=\frac{1}{2}M_n\bm{v}_n^2 + w_n$, where $M_n$ and $\bm{v}_n$ are atomic masses and velocities, and $w_n$ are suitably defined atomic potential energies, \emph{vide infra}. When this is done, the energy flux can be written as 
%the energy flux can be expressed in terms of the atomic positions, $\bm{r}_n$, velocities, $\bm{v}_n$, and energies, $\epsilon_n=\frac{1}{2}M_n\bm{v}_n^2 + w_n$ ($w_n$ is the atomic potential energy) \DT{\cite{Baroni2018}}, by 
% \DT{defined by an atomic partition of the total potential energy s.t. $E=\sum_s E_s$}). \RB{We stress that many different definitions of the atomic energy can produce the same description of the system. For example imagine a shift of the zero of the energy of a single atomic specie or a different partition of the pair potential energy between two interacting atoms.} As in \cite{Baroni2018,Kubo,KuboRyogo,OnsagerI,OnsagerII,DeGroot,Irving-Kirwood} we end up with\RCnote{how is the atomic potential energy defined?} \DTnote{local energies are not uniquely defined even in classical MD }:
\begin{equation}\label{eq:J^e}
    \bm{J}^e(t)=\frac{1}{V} \sum_n \left[  \bm{v}_n \epsilon_n  - \sum_m {(\bm{r}_n - \bm{r}_m)\frac{\partial w_m}{\partial \bm{r}_n}\cdot \bm{v}_n}\right],
\end{equation}
% \RCnote{use better indices, t should not be confused with time } %\DTnote{giusto changed $t$ with $p$ in all the paper} 
where $\bm{r}_n$ are atomic positions and $n$ and $m$ run over all the atoms in the system \cite{Helfand1960,Ercole2016,Baroni2018}. In the case of pair-wise interactions, for instance, it can be assumed that $w_n=\frac{1}{2}\sum_{m\ne n}w(|\bm r_m-\bm r_n|)$. For a general many-body interaction, a similar partition of the total energy into local contributions is also possible.
%, basically expressing the additivity of the energy. 
In a quantum-mechanical setting, it is not possible to uniquely define the atomic energies appearing in Eq. \eqref{eq:J^e}, and the total energy of a system can at most be expressed in terms of an energy \emph{density}, which is also ill-defined. For instance, the electrostatic energy of a continuous charge-density distribution can be expressed as either one half the integral of the density times the potential, or of $\frac{1}{8\pi}$ the squared modulus of the field; by the same token, the kinetic energy of a quantum particle can be expressed as the integral of the squared modulus of the gradient of its wave-function, or of the negative of the product of the wave-function and its Laplacian. For this reason,
% are ill-defined in any quantum-mechanical setting, 
it has long been feared that no quantum-mechanical expressions for the heat conductivity could be obtained from first principles \cite{Stackhouse2010b}. Actually, although not generally fully appreciated, this same problem arises with classical force fields as well, because classical atomic energies themselves are ill-defined. In the example of pair-wise interactions any different partition of the interaction energy of the $nm$ pair into individual atomic contributions would be equally acceptable and, yet, would lead to a different expression for the energy flux \cite{Ercole2016}.

This long-standing problem was solved for good only recently with the introduction of a \emph{gauge invariance} principle for the transport coefficients \cite{Marcolongo2016,Ercole2016,Grasselli2021}, as explained in the following subsections. 

\subsection{Gauge invariance} \label{sec:gauge}
In order to introduce, and understand, the recently discovered \emph{gauge} and \emph{convective} invariance principles for the transport coefficients, it is useful to define the concept of \emph{diffusive} flux. A flux is said to be diffusive if its GK integral, as defined in Eq. \eqref{GKeq}, is different from zero; the flux is said to be non-diffusive otherwise. Gauge invariance states that the addition of any linear combination of \textit{non-diffusive} fluxes to a \textit{diffusive} one does not affect the value of the conductivity calculated with the GK formula, Eq. \eqref{GKeq}.
%\RB{From a more fundamental point of view, non diffusive energetic fluxes can be seen as fluxes arising from different arrangements of contributions to the total energy between different parts of the system. Let's take the continuity equation for the energy density field $e$ and the energy density current $j$ vector field:}
%\begin{equation}    \nabla\cdot \bm j + \dot e = 0 \end{equation}
 This principle got this name because it results from a kind of gauge invariance of conserved densities, according to which any such density is only defined up to the divergence of a bounded vector field. This is so because the volume integral of such a divergence is irrelevant in the thermodynamic limit, and, thus, does not contribute to the value of the conserved quantity.
%\RBnote{my proposal:}\\
% \RB{In the equation above, the energy density $e$ is not unique because, as in the pair potential case, for example, we could decide to assign the interaction energy entirely to a part of the system and not to split it into two equal parts, and the physics would not change. As explained in \cite{Marcolongo2016,Ercole2016,Baroni2018,Marcolongo2021,grasselli2021invariance}, this is equivalent to adding to $e$ a divergence $\nabla\cdot \bm p$ of a bounded vector field $\bm p(t)$.}
This divergence would, in turn, result in the addition of a non-diffusive term to the flux of the conserved quantity, thus not affecting the value of the transport coefficient. 

\subsection{\label{sec:MultiComp} Convective invariance}
%\RB{Convective invariance aims to solve a different issue than the energy partition problem. Suppose for example that we change the zero of the energy for a single atomic specie in my system. That is adding a constant to the atomic potential energy. If the center of mass of all the atoms of the specie is a diffusive quantity, this shift adds a diffusive flux to the energy current \eqref{eq:J^e}, but the physics of the system is unchanged. Thus, it is natural to expect that the thermal conductivity remain the same, as it will be clear after explaining the theory behind transport in multicomponent systems, describing the phenomenological interaction between all the conserved quantities.}

 In general, a system made of $M$ atomic species (an $M$-\emph{component system}) has $M+4$ conserved quantities (the number of atoms of each species, the energy, and the three components of the momentum). The energy and atomic-number currents are vector quantities, whereas the momentum currents are $3\times 3$ (stress) tensors, which do not couple with the former in a rotationally invariant system. The total momentum is not only a conserved quantity by itself, but is also a linear combination of the volume integral of the atomic-number currents (\emph{atomic-number fluxes}). This reduces the number of independent mass fluxes from $M$ to $M-1$. We conclude that,
when dealing with an $M$-component system, the conserved quantities relevant to heat transport are the total energy and the total numbers (or masses) of each one of the $M-1$ independent atomic components, which, in the linear regime, are related to each other by Onsager's phenomenological relations:
\begin{equation}\label{eq:Onsager}
    \bm{J}^i = \sum_{j=0}^{M-1}\Lambda^{ij} \bm F^j,
\end{equation}
where $\bm F^j$ is the thermodynamic force associated to the $j-th$ conserved quantity being transported. In Eq. \eqref{eq:Onsager} the energy flux is identified as the zero-th term, the remaining $M-1$ fluxes being any linearly independent combinations of the mass fluxes, and the $\Lambda$ coefficients are expressed by the GK integrals:
%. As the total-mass flux is the total momentum, which is also a constant of motion, the number of relevant conserved fluxes is reduced from $M+1$ to $M$: energy, which we label as the zeroth, and $M- 1$ convective fluxes, which can be identified with any independent linear combinations of the molecular mass or number fluxes (\emph{e.g.} just any $M-1$ of them). Calling $\Lambda^{ij}$ the cross GK integrals
\begin{equation}\label{eq:Lambda}
    \Lambda^{ij} =  \frac{V}{3k_B}\int_0^{\infty} \langle \bm{J}^i(\Gamma_t) \cdot \bm{J}^j(\Gamma_0) \rangle dt.
\end{equation}
%where $i,j= 0 \dots M-1 $, \DTchange{transport is described in the linear regime in terms of the Onsager relations}{the flux $J^i$ satisfies the \RB{phenomenological} Onsager relation in the linear regime}:
%\begin{equation}\label{eq:Onsager}
%    \bm{J}^i = \sum_{j=0}^{M-1}\Lambda^{ij} \bm F^j,
%\end{equation}
% where $\bm F^j$ is the thermodynamic force relative to the $j-th$ conserved quantity being transported. 
In the multi-component case, the heat conductivity is defined as the ratio between the energy current and the negative of the temperature gradient, \emph{when all the mass currents vanish}. With some simple algebra, we arrive at the expression
%With these definitions a general relation for the heat conductivity can be \DTchange{defined as}{best achieved by partitioning the $\Lambda$ matrix into a $1 \times 1$ energy block, $\Lambda^{00}$$ and a $(M-1) \times (M-1)$ \emph{mass block}, $\lambda_{M-1}$, and by performing an inversion} 
\cite{Bertossa2019}:
\begin{equation}\label{eq:kappa_multi}
    \kappa = \frac{1}{T^2} \left[ \Lambda^{00}-\sum_{i,j=1}^{M-1}\Lambda^{0i}(\Lambda_{M-1}^{-1})^{ij}\Lambda^{j0} \right],
\end{equation}
where $\Lambda_{M-1}^{-1}$ is the inverse of the $(M-1)\times(M-1)$ mass block of the Onsager matrix. The expression in square brackets in Eq. \eqref{eq:kappa_multi} is called the \emph{Schur complement} of the mass block in the Onsager matrix, and is nothing but the inverse of the $00$ element of the inverse Onsager matrix.

By combining the definition of $\Lambda$ with \cref{eq:kappa_multi}, one can demonstrate by a straightforward substitution that the heat conductivity is invariant with respect to the addition of any linear combination of mass fluxes to the energy flux: $\bm J^0 \to \bm J^0 + \sum_{i=1} ^{M-1} c^i \bm J^i$. This is the transformation the energy flux undergoes
when the energies of all the atoms of the same chemical species are shifted by the same amount, such as it occurs, \emph{e.g.}, when passing from an all-electron to a pseudo-potential representation of the electronic structure, or when changing pseudo-potentials. This property has been called \emph{convective invariance} \citep{Bertossa2019}

%\RCnote{A clearer exposition would be welcome. Rather than saying strictly speaking etc.. go directly to the statement! One should separate the concept of heat conductivity and its relation to the appropriate fluxes and that of the time series to optimize data analysis}\DTnote{Not sure what Roberto mean, I think we should say that the multicomponent is not necessary from a theoretical point of view but very powerful for the practical data analysis }

%\RB{It turns out that \eqref{eq:kappa_multi} is very useful also when we are dealing with system where there is no diffusion for any of the mass currents $\bm J^i$, where the previously stated gauge invariance principle and \eqref{GKeq} would be sufficient to compute a physically meaningful thermal conductivity. This is the case of}

Molecular fluids, such as undissociated water, deserve a special comment. In this case, one demonstrates that, as the atoms in each molecule do not diffuse relative to the center of mass of the molecule, all the independent atomic mass/number fluxes are non-diffusive. Therefore, energy can be assumed to be the only conserved flux relevant to heat transport, as it is the case for strictly one-component fluids \cite{Marcolongo2016}.

Notwithstanding gauge and convective invariance, the statistical noise affecting the estimate of the heat conductivity does depend on the 
energy flux of the non-diffusing components that are added to the diffusive energy flux. Gauge invariance can then be leveraged to tune the optimal linear combination of non-diffusive fluxes to minimize the statistical error on the heat conductivity. In order to achieve this goal, it is expedient to consider the transport coefficient as the zero-frequency value of $S(\omega)$, the flux power spectrum, which is given, in the multi-component case, by:
%
%\RBcancel{Strictly speaking, this principle is not of much use for} (undissociated) molecular fluids, such as water at ambient conditions \cite{Marcolongo2016}, or solids.\RBcancel{, because it can be demonstrated that mass fluxes are non-diffusive in this case} However,\RBcancel{ convective and gauge transformations are powerful tools for the data analysis of the flux time series from which transport coefficients are computed, because,} while conductivities are not affected by \RBchange{these transformations}{using \eqref{eq:kappa_multi} with non diffusive mass fluxes in place of \eqref{GKeq} with the energy current alone}, the statistical error affecting \RBchange{them}{$\kappa$} do depend on the choice, so that these transformations can be utilized to optimize data analysis \cite{Bertossa2019,Marcolongo2021}. The expression for the heat conductivity, Eq. \eqref{eq:kappa_multi}, can conveniently be obtained from the \textit{cross power spectrum},
\begin{equation}\label{eq:S_omega}
     S(\omega) = \frac{V}{2k_BT^2}\frac{1}{[\bar S^{-1}(\omega)]^{00}},
\end{equation}
where $[\bar S^{-1}(\omega)]^{00}$ is the $00$ element of the inverse of the matrix defined by:
\begin{equation}\label{eq:Sij}
    \bar{S}^{ij}(\omega) = \frac{1}{3}\int_{-\infty}^{\infty} \langle \bm{J}^i(\Gamma_t) \cdot \bm{J}^j(\Gamma_0) \rangle e^{-i\omega t} dt.
\end{equation}

%of which $\kappa$ is simply the $\omega=0$ value 
In molecular fluids, 
%\SBcancel{where} 
all mass fluxes are non diffusive \cite{Marcolongo2016} and energy is the only conserved quantity relevant to heat transport. Therefore, we actually have $S(0)=\frac{V}{2k_BT^2}\bar S^{00}(0)$ and, strictly speaking, no multi-component analysis would be needed. However, data analysis is greatly facilitated when the power spectrum is as smooth as possible (to be precise, when the number of inverse Fourier coefficients of the logarithm of the spectrum are as few as possible \cite{Ercole2017}). For this reason, it may be convenient to complement the diffusive energy flux with a number of non-diffusive ones, which, while not altering the value of the spectrum in Eq. \eqref{eq:S_omega} at $\omega=0$, decrease the total power, thus easing data analysis \cite{Bertossa2019,Baroni2018,Grasselli2021,Marcolongo2020}.
%, it is convenient to use \cref{eq:S_omega} since the power of $S(\omega)$ is in general smaller than that of $\bar S^{00}(\omega)$ at $\omega\ne 0$, thus making the data analysis techniques more effective.

\subsection{\label{sec:abThermo} The MUB DFT adiabatic energy flux}

Gauge invariance solves the problem of the alleged indeterminacy of the quantum-mechanical adiabatic energy flux, thus providing a rigorous derivation of its expression within DFT, without introducing any ad-hoc ingredients \cite{Marcolongo2016}. Within the local density (LDA) and generalized gradient (GGA) approximations of DFT, the MUB expression for the DFT energy flux \cite{Marcolongo2016,Marcolongo2021} is:
\begin{equation}
 \bm{J}^{MUB}  =\bm{J}^{KS} +\bm{J}^{H}+ \bm{J}^{0}+ \bm{J}^{n}  + \bm{J}^{XC}, \label{eq:JAris}
\end{equation}
where
\allowdisplaybreaks
\begin{align}
    \bm{J}^{KS} &=\sum_{v} \left (\langle\varphi_{v}| \bm{\hat r}\hat{H}^{KS}| \dot{\varphi}_{v}\rangle + \varepsilon_v \langle\dot{\varphi}_{v}| \bm{\hat r} | \varphi_{v}\rangle \right), \nonumber \\
    \bm{J}^{0}  &= \sum_{n \bm L} \sum_v \left\langle \varphi_{v} \left|(\bm{\hat r}-{\bm r}_n - {\bm L}) \left(\bm{v}_{n} \cdot \nabla_{n \bm L} \hat{v}^0  \right)\right|\varphi_{v} \right\rangle , \quad \nonumber \\
    \bm{J}^{n} &= \sum_{n}\left [ \bm{v}_n  e^0_n -\sum_{\bm{L}\neq 0} \bm{L}\bigl(\bm{v}_{n} \cdot \nabla_{n \bm L }w_n^Z \Bigr) \right .  \label{eq:J_H}\\
    & \quad\quad  \left . +\sum_{m \neq n}\sum_{\bm L}(\bm{r}_n-\bm{r}_m -\bm{L})\left(\bm{v}_{m} \cdot \nabla_{m \bm L}w_n^Z\right) \right ] \nonumber \\
    %\bm{J}^{n}  & =\sum_{s} \left [ \bm{V}_{s} e^0_s + \sum_{t\ne s}{}^' (\bm{R}_{s}- \bm{R}_{t}) \left(\bm{V}_{t} \cdot \frac{\partial w_s}{\partial \bm{R}_{t}} \right)\right ], \label{eq:J_n} \\
   \bm{J}^{H} &=\frac{1}{4\pi e^2} \int\dot{v}^{H}(\bm{r}) \nabla v^{H}(\bm{r}) d\bm{r}, \nonumber \\
   \bm{J}^{XC} &=  \begin{cases}
        0 & \text{(LDA)} \\ -\int n(\bm{r})\dot{n}(\bm{r}) \bm \partial\epsilon^{\text{GGA}} (\bm{r}) d\bm{r} & \text{(GGA)},
    \end{cases} \nonumber
\end{align}
%where $\bm{r}_n$, $\bm {v}_n$ are the ionic positions, velocities, %respectively, and $w_n^Z$ is the electrostatic energies so defined:
%\begin{equation}
%     w_s^Z = \frac{e^2}{2} \sum_{m\neq n}\sum_{\bm %L}\frac{Z_mZ_n}{\vert \bm{r}_n-\bm{r}_m-\bm{L}\vert} + %\frac{1}{2}e^2 Z_n^2\sum_{ L\neq 0}\frac{1}{L}
%\end{equation}
%where $Z_n$ are the ionic charges; $\bm L$ is a lattice vector, $\nabla=\partial/\partial\bm{r}$ and $\nabla_{m \bm L}=\partial/\partial\bm{r}_{m \bm L}$ represent, respectively, the gradients with respect to the argument of the function and to the atom at position $ \bm{r}_m + \bm L$;
where $\bm{r}_n$, $\bm {v}_n$, and $w^{Z}_n=1/2\sum_{m\neq n}^{'}(Z_m Z_n/\vert \bm{r}_n-\bm{r}_m\vert)$ are ionic positions, velocities, and electrostatic energies, respectively, $Z_n$ are ionic charges, and $\sum^{'}$ includes all the atoms in the cell and their periodic images; 
%\RCnote{why the prime in the sum when we say explicitly that $m \neq n$ ?} \DTnote{$m$, $n$ are atom in the computational cell the primed sum extend %the sum to all atoms even outside the box to the images}
$\hat{H}^{KS}$ is the instantaneous Kohn–Sham (KS) Hamiltonian, $\varphi_{\nu}$ and $\varepsilon_{\nu}$ are the occupied eigenfunctions and corresponding eigenvalues, and $\rho(\bm{r})=\sum_{\nu}\vert \varphi_{\nu}(\bm{r})\vert^2$ is the ground-state electron-density distribution; $v_H$, $v_{XC}$ are Hartree and exchange-correlation (XC) potentials; $\bm L$ is a lattice vector, $\nabla=\partial/\partial\bm{r}$ and $\nabla_{m \bm L}=\partial/\partial\bm{r}_{m \bm L}$ represent, respectively, the gradients with respect to the space position $\bm{r}$ and with respect to the atom position at $ \bm{r}_m + \bm L$ (that is an image if $\bm L\neq 0$); $ \hat{\nu}_0$ represents the (possibly non-local) ionic (pseudo-) potential acting on the electrons; LDA and GGA indicate the local-density \cite{KohnSham} and generalized-gradient \cite{PBE} approximations for the XC energy functional and $\partial \epsilon^{\text{GGA}}$ is the derivative of the GGA XC local energy per particle with respect to density gradients. 
% $\bar{\bm{\psi}}_{\nu}$ is the projection of $\bm{r}\vert \psi_{\nu} \rangle$ over the empty-state manifold, $\bar{\bm{\psi}}_{\nu} = \hat{P}_c[\hat{H}_{KS},\bm{r}]\vert\psi_{\nu}\rangle$, where $\hat{P}_c=1-\sum_{\nu} \vert \psi_{\nu}\rangle \langle\psi_{\nu}\vert$ and can be computed from standard density-functional perturbation theory \cite{BaroniDFPT}. 
All the terms in \cref{eq:JAris} are well defined under periodic boundary conditions (PBC) \cite{Marcolongo2016}. Only the 
expression of $\bm{J}^{KS}$ depends on the choice of the arbitrary zero of the one-electron energy levels. A shift of this zero by $\Delta\epsilon$ results in a KS energy flux shifted by $\Delta\epsilon \bm{J}^{\rho}$, $\bm{J}^{\rho}$ being the adiabatic electronic flux \cite{Thouless}, $ \bm{J}^{\rho}=2\sum_{v}\langle\dot{\varphi}_{v}| \bm{\hat r} | \varphi_{v}\rangle$ (the factor 2 accounts for spin degeneracy in a singlet state), which is also well defined within PBC. The adiabatic electronic flux is non-diffusive, being the difference between the total-charge flux, which is by definition non-diffusive in insulators \cite{grasselli2019}, and its ionic component, non-diffusive in mono-atomic and molecular systems, because of momentum conservation and the condition that molecular bonds do not break \cite{Marcolongo2016, Bertossa2019}. Therefore, $\bm{J}^{\rho}$ does not contribute to the heat conductivity, thus lifting this further apparent indeterminacy of the transport coefficient derived from the MUB energy flux.

\section{\label{sec:deepMD}{Deep Potential} model}
To speed up equilibrium MD simulations, we trained a DNN model according to the DP framework \cite{NIPS2018_7696}. Consider a system of $N$ atoms, whose configurations are represented by the set of atomic positions, $\bm{r} = \left\{ \bm{r}_1, \bm{r}_2,\dots ,  \bm{r}_N\right\} \in \mathbb{R}^{3N}$. For each atom, $n$, we consider only the neighbours, $\{q\}$,  such that $r_{qn}<r_c$, where $r_{qn}$ 
% is the distance between the $s$-th and $t$-th atoms 
is the modulus of the vector $\bm{r}_{qn} = \left[x_{qn},y_{qn},z_{qn}\right]\doteq \bm{r}_{q} - \bm{r}_{n}$, 
and $r_c$ is a pre-defined cut-off radius. Denoting with $N_n$ the number of neighbours of $n$ within the cutoff radius, we define the \emph{local environment matrices} $\tilde{R}_n \in \mathbb{R}^{N_n\times 4}$  to encode the local environment:
\begin{align} \label{eq:Rtilde}
\tilde{R}_n  &=
\begin{bmatrix}
\frac{\sigma(r_{1n})}{r_{1n}} & \frac{\sigma(r_{1n})x_{1n}}{r_{1n}^2} & \frac{\sigma(r_{1n})y_{1n}}{r_{1n}^2} & \frac{\sigma(r_{1n})z_{1n}}{r_{1n}^2} \\
\frac{\sigma(r_{2n})}{r_{2n}} & \frac{\sigma(r_{2n})x_{2n}}{r_{2n}^2} & \frac{\sigma(r_{2n})y_{2n}}{r_{2n}^2} & \frac{\sigma(r_{2n})z_{2n}}{r_{2n}^2} \\
\vdots & \vdots & \vdots & \vdots 
\end{bmatrix},
\end{align}
where $\sigma(r_{qn})$ is a smoothing function (see \cref{sec:derivatives}). Then, symmetry-preserving descriptors (extensive details in \cite{NIPS2018_7696}) are constructed and fed to the DNN, which returns the local energy contribution $w_n$ in output. We denote by $\bm W$ the full set of parameters that define the total potential energy, $E$.
Thus, as illustrated in Ref. \onlinecite{NIPS2018_7696}, the extensive property of $E$ is ensured by its decomposition into ``atomic contributions":
\begin{equation}
    E^{\bm W} (\{\tilde{R}\}) = \sum_n w^{\bm W_{\alpha _n}}(\tilde{R}_n)\equiv \sum_n w_n
\end{equation}
where $\alpha _n$ denotes the chemical species of atom $n$. We use the notation $(\dots)^{\bm W_{\alpha _n}}$ to indicate that the parameters used to represent the ``atomic energy'', $w_n$, only depend on the chemical species $\alpha_n$ of the $n$-th atom. Being $w_n$ a well defined and easy to compute function of the atomic positions, the atomic forces and their breakup into individual atomic contributions, $\frac{\partial w_m}{\partial\bm{r}_n}$ (needed in the definition of the energy flux in \cref{GKeq}), can be easily computed as the gradients of $E$ and $w_n$, respectively. In particular, the computation of the latter can be divided into two contributions by applying the chain rule:
\begin{align}
\nabla_{\bm{r}_n} w_m &= \frac{\partial w_m}{\partial \bm{r}_n} = \sum_{i,j}
 \frac{\partial w_m}{\partial \tilde{R}_m^{ij}}  \frac{\partial \tilde{R}_m^{ij}}{\partial \bm{r}_n}\label{eq:1der}
\end{align}
where $i,j$ identifies an element of the matrix $\tilde{R}_m$. 
The first terms can be easily computed with TensorFlow \cite{tensorflow2015-whitepaper}, while the second must be handled separately and coded explicitly \cite{NIPS2018_7696,Linfeng2018}. A more detailed description of the calculation can be found in \cref{sec:derivatives}. The local energy and its derivatives are the key elements in the computation of the energy flux, \cref{eq:J^e}.
The parameters of the model are determined by minimizing the loss function:
\begin{equation}\label{eq:Loss}
    L=p_E \Delta E^2 + \frac{p_f}{3N}\sum_n \Delta \bm F_{n}^2 
\end{equation}
where $\Delta E^2 $ and $\Delta \bm F_n^2 $ are the squared deviations of the potential energy and atomic forces, respectively, between the reference DFT model and the DNN predictions. The two prefactors, $p_E$ and $p_f$, are needed to optimize the training efficiency and to account for the difference in the physical dimensions of energies and forces.

%\subsection{\label{sec:invariances} Invariance principles and Neural Networks}
%In recent years, two powerful invariances of the thermal transport coefficient has been discovered: the \textit{gauge invariance}\cite{Ercole2016} and the \textit{convective invariance} \cite{Bertossa2019}. Both principles tell us that the heat conductivity is largely independent on the details of the microscopic definition of the energy densities and fluxes from which it is derived. This idea allowed Marcolongo, Umari and Baroni \cite{Marcolongo2016} to build a DFT adiabatic energy flux despite the absence of properly defined DFT local energies. This ideas turn to be crucial even using neural network where the energy distribution into local contribution is rather arbitrary only constrained by the total energy of the system. 

We remark that gauge invariance is instrumental in ensuring the uniqueness of the heat conductivity in a DNN framework. In fact, the roughness of the loss-function landscape implies that equally good representations of the potential-energy surface and atomic forces may be reached with very different representations of the atomic contributions to the total energy. Gauge invariance implies that, if the total energies resulting from two different local representations were identical, the resulting transport coefficients would also be identical, thus making them in practice dependent on the overall accuracy of the DNN model, but not on the details of its local representation.

%Since the loss function's landscape has many equivalent minima, models trained on the same set, same number of parameters but different initialization will generate different local energy contributions.  In general, neural network models trained over the same data set but with different initial values of the weights, generates different, but very close, total potential energies but generate completely different local atomic energies. The gauge invariance ensure us that the thermal transport coefficient is the same, thus elevating the \textit{gauge invariance} at key theoretical component in the computation of thermal transport .

\section{Results} \label{sec:results}
\subsection{Ab initio Molecular Dynamics} \label{sec:abinitio_results}
% \section{Ab-initio results}

We performed four \ai MD simulations of water, corresponding to different temperatures and phases, using the PBE functional approximation of DFT, the plane-wave pseudopotential method, and periodic boundary conditions. Hamann-Schlüter-Chiang-Vanderbilt (HSCV) norm-conserving pseudopotentials \cite{Hamann2013} were used with a kinetic-energy cutoff of 85 Ry. All the simulations were performed with the Car-Parrinello extended-Langrangian method \cite{Car:1985} using the \texttt{cp.x} component of \qe\ \cite{Giannozzi_2009,Giannozzi_2017,QE3} and setting the fictitious electronic mass to $25$ physical masses and the timestep to $dt=0.073~$fs.  Liquid water simulations were done with $125$ water molecules inside a cubic computational box of side $l=15.52~$\AA, hexagonal ice-Ih simulations used $128$ water molecules inside an orthogonal cell, with sides: $l_1=18.084~$\AA, $l_2=15.664~$\AA~and $l_3=14.724~$\AA. It is known that within the PBE XC functional approximation, liquid water
exhibits enhanced short-range order \cite{Grossman2004,Schwegler2004} and a melting temperature that is more than $100$~K higher than in experiment \cite{Sit2005,Soohaeng2009}, while solid ice has higher density than liquid water at coexistence. In order to compensate for this shortfall, it is customary to offset the simulation conditions by increasing the temperature by $\approx 100$~K. We performed simulations of the liquid at three temperatures ($521~$K, $431~$K and $409~$K), and of ice 
% In particular, taking into account that PBE water reproduce the behavior of experimental water at $100~$K lower  and that the freezing point of PBE water is estimate to be around $400K$ \cite{Sit2005,Soohaeng2009}  three liquid systems at $521~$K, $431~$K and $409~$K and an 
in the hexagonal Ih structure at $260~$K. Each simulation was $100~$ps long. Then, using the \qeheat\ \cite{Marcolongo2021} code, we computed 
the MUB flux every $3.1~$fs. The statistical noise affecting the estimates of the GK integrals is larger when the spectral power of the flux time series is larger. Because of gauge invariance, different representations of the energy current may carry a very different spectral power, and still yield the same conductivity, which is the zero-frequency limit of the flux power spectrum. The MUB energy flux turns out to carry an impractically large spectral power, which can be tamed to some extent by leveraging gauge and convective invariance. Gauge invariance is first exploited by the \emph{velocity renormalization} technique of Ref. \onlinecite{Marcolongo2020}. In a nutshell, it can be demonstrated that subtracting to each atomic velocity the average velocity of all the atoms of the same chemical species, results in a current with a much reduced spectral weight but the same conductivity. Further spectral weight can be subtracted by adding to the resulting effective flux any linear combination of non-diffusive fluxes. This can be effectively done by treating the (possibly renormalized) energy current as one component of an $M$-component system, where all the other currents are non-diffusive ones \cite{Bertossa2019}. Here, we choose $M=2$ and take the electronic adiabatic current as the auxiliary non-diffusive one. 
%
%Evaluating $\kappa$ from the MUB flux is not straightforward, as the  is made noisier the larger the spectral power 
%
%since the bare GK integral contains a great amount of noise that would lead to an incorrect value of $\kappa$. A lot of work has been done in developing new data-analysis \cite{Ercole2017,Bertossa2019} and noise reduction techniques \cite{Marcolongo2020}. In particular it has been shown \cite{Marcolongo2020,Bertossa2019} that the analysis can be highly improved by decorrelating the MUB flux from the adiabatic electronic flux and the mass flux. We can, thus, consider our water system composed of $M=3$ fluxes ($M=0$ is the energy flux) and apply the multicomponent approach, \cref{eq:kappa_multi}. An equivalent approach for decorrelating the mass flux is renormalize the velocity, subtracting to every atom speed the velocity of the center of mass of the respective species \cite{Marcolongo2020}. Then, only two fluxes are left: the energy and electronic flux. The thermal transport coefficient is computed as $\omega=0$ value of the power spectrum in \cref{eq:S_omega}. 
In all cases, the transport coefficients are obtained from the \emph{cepstral analysis} \cite{Ercole2017,Bertossa2019} of the power spectrum of the relevant currents, using the \sportran\ \cite{SporTran} code.

\begin{figure}[tb]
    \centering
    \includegraphics[width=\linewidth]{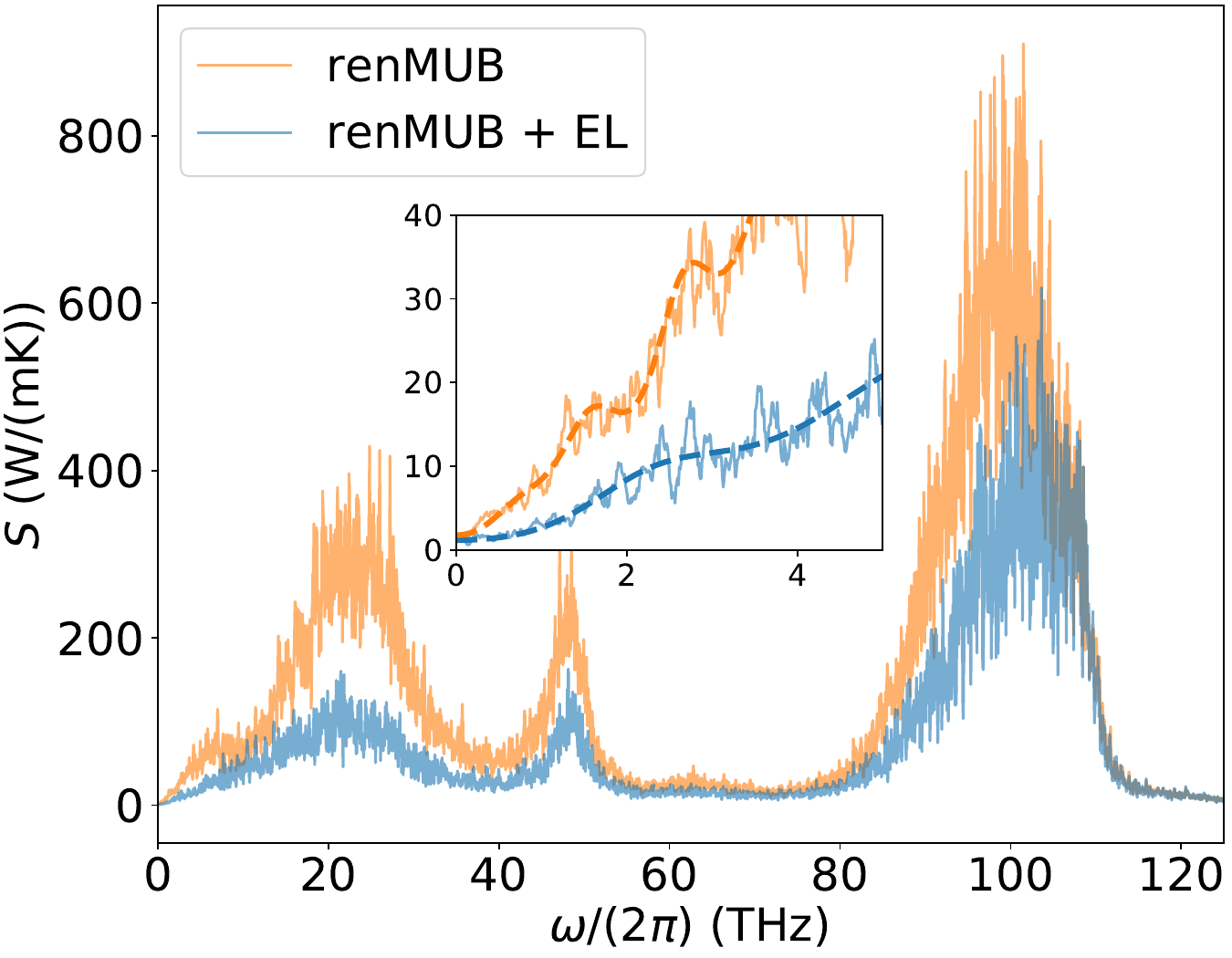}
    \caption{Comparison of the (window-filtered) spectrum of the velocity renormalized MUB flux (orange) and of the velocity renormalized MUB flux decorrelated with the adiabatic electronic flux (blue). Both spectrum are filtered with a moving average of 0.1 THz. The renormalized MUB flux has a higher power but close to zero the two spectra converge to the same value. The two dashed lines in the inset represent the cepstral filters of the power spectra. }
    \label{fig:Spectrum_renEL}
\end{figure}

\cref{fig:Spectrum_renEL} displays the (window-filtered) power spectrum of the MUB flux from one of our Car-Parrinello MD simulations of liquid water at an average temperature of $431~$K, using renormalized velocities (orange line), and further removing the contribution of the adiabatic electron current from the energy flux (blue line). In the inset we see that the two spectra converge to the same value when $\omega=0$. The decorrelation decreases the power of the spectrum and flattens the spectrum near $\omega = 0$ facilitating data analysis by reducing the number of the required cepstral coefficients.

\subsection{\label{Water_properties}DPMD benchmark against GGA results}

\begin{figure}[htb]
     \centering
     \includegraphics[width=\linewidth]{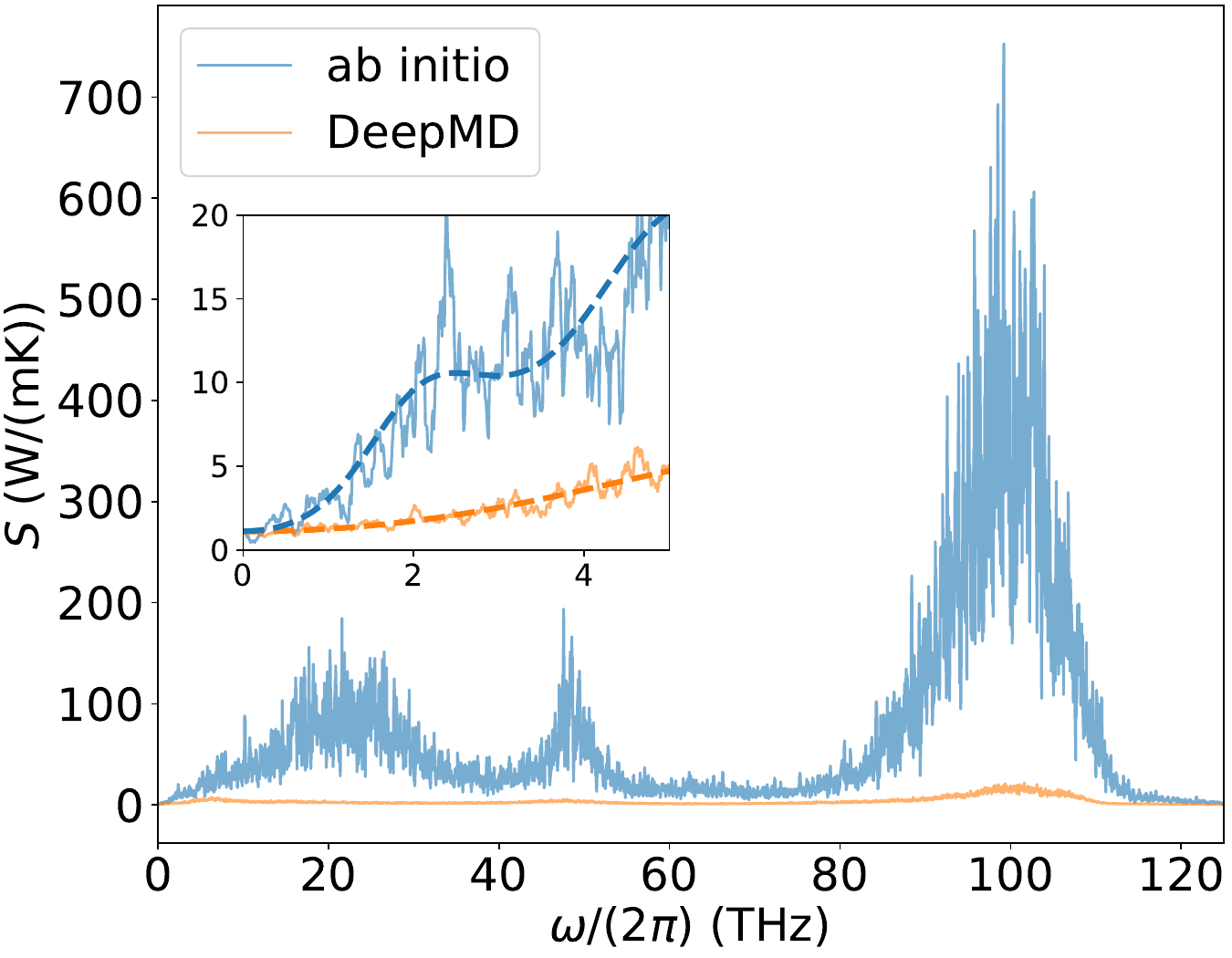}
     \caption{Power spectrum of a water simulation. The orange line is obtained from $ 360~$ps of DPMD simulation of a periodic cubic cell containing $125$ water molecules at $ 407~$K. The blue line is obtained from an \emph{ab initio} MD simulation of $125$ water molecules with the same cubic box and an average temperature of $ 409~$K. Both spectrum are filtered with a moving average of 0.1 THz. The dashed lines in the inset represent the cepstral-filtered spectra. Even though the two spectra have very different intensities the values at zero frequency are the same. }
     \label{fig:pspectrum}
 \end{figure}

In order to appraise the ability of DP models to accurately describe heat transport phenomena, we have generated one such model, by training it on a set of DFT-PBE data extracted from Car-Parrinello trajectories at different temperatures in the [400K -- 1000K] temperature range. The loss function in \cref{eq:Loss} was optimized with the Adam stochastic gradient descent method \cite{Kingma2014}. The details of the training protocol are given in \cref{sec:NeuralNetworkTraining}. The generated DNN potential was then used to run equilibrium MD simulations of water at the same conditions explored in the previous subsection by \ai techniques.
% First of all, we compare the result computed from \ai molecular dynamics at PBE gga level of theory and a neural network potential  trained over the same data. In order to obtain the training set of data we selected snapshot from simulations of water at different temperatures, in the range $[400-1000]$K 
%$409$K, $431$K, $521$K, $800$K and $1000$K. 
% Then, we optimize the loss function in \cref{eq:Loss} through the Adam stochastic gradient descent method \cite{Kingma2014}. We stopped the training after $1500000$ training steps, for more details on the parameters of the training and on the validation of the model see \cref{sec:NeuralNetworkTraining}.
One of the resulting energy-flux power spectra is displayed in \cref{fig:pspectrum} (orange), together with the corresponding \ai\ spectrum (blue). 
The thermal conductivities corresponding to the two spectra are obtained as before through cepstral analysis. Notice that, in spite of the much larger weight of the \emph{ab initio} spectrum relative to that of the DNN model, the two spectra have the same low-frequency limit, indicating that the two simulations predict the same conductivity within statistical errors. The difference between the two spectra stems much more from the different local representations of the potential energy than from a different dynamics. The latter is, in fact, very well mimicked by the DNN potential, which gives forces in close agreement with those of the \emph{ab initio} model (see \cref{Benchmark}).
%We, then, employed the neural network potential, with a timestep of $0.2~$fs, to simulate water at the same conditions explored in the previous section by \ai simulations.
% We can proceed to compute the heat flux of water via the Green-Kubo formulation, \cref{eq:J^e}, and compute the thermal transport coefficient through the analysis of the spectrum showed in \cref{sec:MultiComp}. \cref{fig:pspectrum} shows the power spectrum both for the \ai (orange) and NN (blue) simulations. Even though the first is even two order of magnitude bigger than the latter, in the insect we can see that both reach the same value at zero. As already said, a more flatten and lower intensity spectrum near zero implies a much easier data analysis.

In \cref{tab:resultkappaPBE} we display the thermal conductivities computed from \ai MD and DPMD for all the simulations that we performed, together with the atomic diffusivities, $D_H$ and $D_O$.
%
%shows the diffusivities of hydrogen, $D_H$, the diffusivities of oxygens, $D_O$, and the thermal transport coefficient $\kappa$ for all the different water systems analised both from \ai and with with the NN.  The diffusivities 
The latter are computed from the $\omega=0$ value of the power spectrum of the velocity:
\begin{align}\label{eq:D_omega}
 \bar{D}_{\alpha}(\omega)  =&  \frac{1}{6N_{\alpha}} \sum_{n}^{N_{\alpha}} \int_{- \infty}^{ \infty} \langle \bm v_{n}(0)\cdot \bm v_{n} (t) \rangle ~ e^{i \omega t} dt
\end{align}
where $\alpha$ represents the atomic species (oxygen and hydrogen here) and $n$ runs over all the atoms of species $\alpha$. The diffusivities are obtained from a block analysis of a $100~$ps long trajectory. The DP model was capable of reproducing accurately the three transport coefficients. In particular, it allowed us to perform longer simulations in order to reduce the statistical uncertainty on $\kappa$. While $\approx 100~$ps long trajectories suffice for errors of about $10$\% in liquid water and of about $20$\% in ice Ih, we found that $\approx360~$ps long trajectories
with the DP model reduced these errors to $5$\% and $8$\%, respectively. These errors could be reduced even further because trajectories lasting tens of ns or more would be possible with DPMD. 

\begin{table*}[htb]
    \centering
    \begin{tabular}{c|c|c|c|c|c}
     & phase   & $T$ & $D_{H}$ & $D_{O}$ & $\kappa$ \\
     &             & K & \AA$^2$/ps & \AA$^2$/ps & W/(mK) \\
        \hline 
        \multirow{4}{*}{DPMD} & liquid  & $516 $ & $1.07 \pm 0.05 $ & $1.08 \pm 0.05$ & $0.99 \pm   0.05 $ \\
        & liquid  & $423 $ & $0.41 \pm 0.02 $ & $0.42 \pm 0.02$ & $1.03 \pm 0.05 $ \\
       & liquid  & $408 $ & $0.29 \pm 0.02$ & $0.32 \pm 0.02$ & $1.11 \pm 0.05$ \\
         & ice Ih & $270$ & - & - &  $1.9 \pm 0.2$ \\
         \hline
        \multirow{4}{*}{\ai} & liquid  & $521 $ & $1.13 \pm 0.05 $ & $1.11 \pm  0.05$ & $ 0.98 \pm 0.19 $ \\
        & liquid  & $431$ & $0.45 \pm 0.03$ & $0.45 \pm 0.03$ & $1.06 \pm 0.11$ \\
       & liquid  & $409$ & $0.325 \pm 0.018$ & $0.29 \pm 0.02$ & $1.12 \pm 0.17$ \\
        & ice Ih  & $260$ & - & - & $1.8 \pm 0.4$ \\
    \end{tabular}
    \caption{Comparison of some properties of water from \ai MD and DPMD simulations based on PBE-DFT. All liquid simulations used $125$ H$_2$O molecules inside a cubic box of side $l=15.52~$\AA. The ice Ih simulations used $128$ H$_2$O molecules inside an orthogonal cell with sides: $l_1=18.084~$\AA, $l_2=15.664~$\AA~and $l_3=14.724~$\AA. $T$ is the mean temperature of the simulations; $D_H$ and $D_O$ are the diffusivities of hydrogen and oxygen, respectively; while $\kappa$ is the thermal transport coefficient. The diffusivities of ice Ih are compatible with zero and are not reported. }
    \label{tab:resultkappaPBE}
\end{table*}

The calculated heat conductivities with DPMD and \ai MD, based on PBE-DFT, agree closely among them, but differ substantially from experiment ($\kappa_{expt} \approx 0.6~$W/(mK) \emph{vs.} $\kappa_{PBE} \approx 1~$W/(mK) for water at near ambient conditions \cite{Ramires1995}), indicating that the distribution of the energy density resulting from the PBE functional adopted here is likely inadequate to accurately describe adiabatic energy transport in water. This prompted us to try more advanced functional approximations, like  
the meta-GGA SCAN framework, to cope with this shortcoming.

% value of $0.606~$W/(mK) for water close to ambient conditions \cite{Ramires1995}, pushing us to look for a functional that can represent better the water properties.
%\cref{fig:kappa_T125} show the values of $\kappa$ for different temperatures and compare the result obtained from classical MD simulation of $125$ water molecules (blue line) with the experimental values (black line).

\section{Extended simulations with a SCAN based deep potential model} \label{sec:SCAN}
\begin{figure}[!h]
    \centering
    \includegraphics[width=\linewidth]{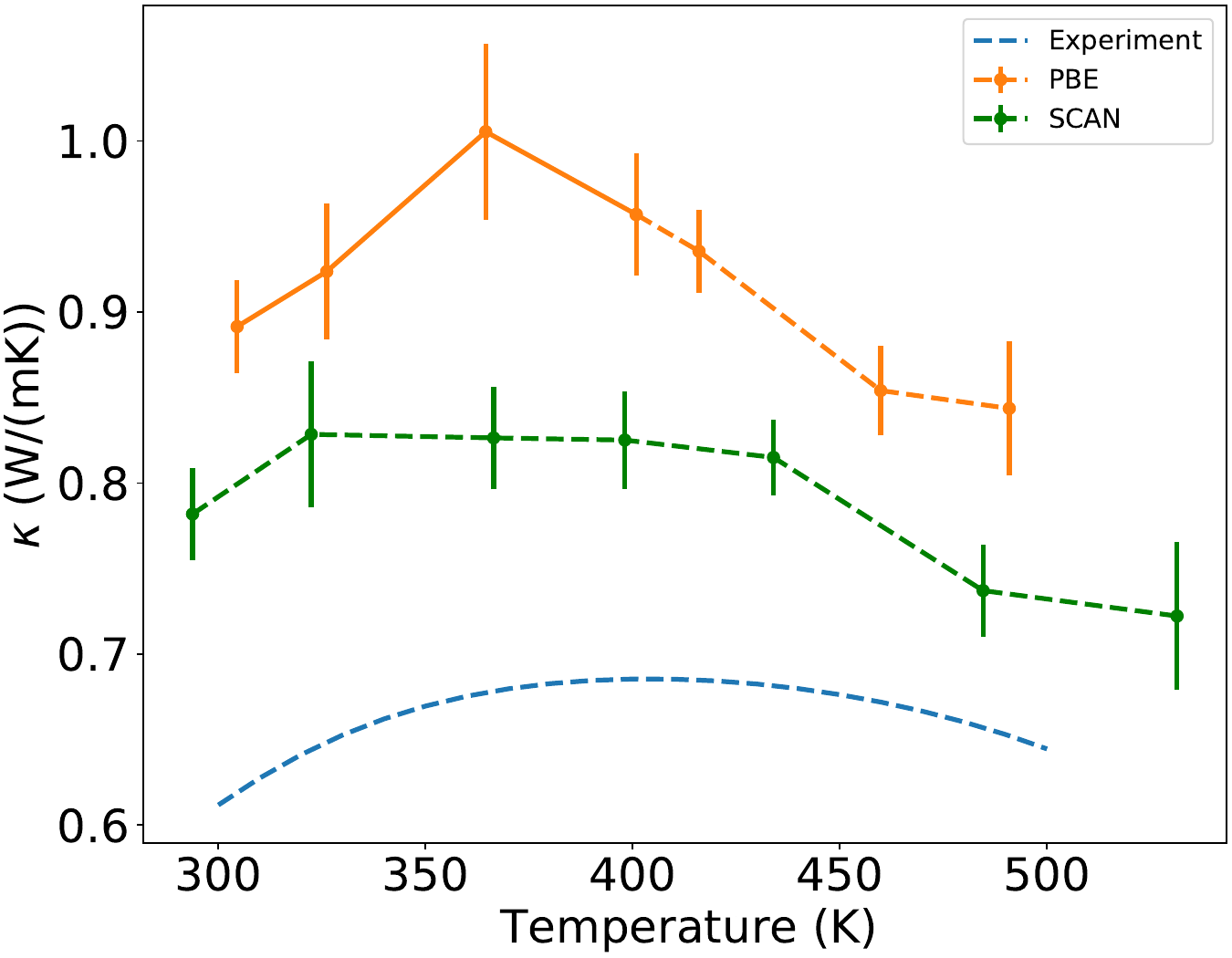}
    \caption{Temperature dependence of the thermal conductivity $\kappa$ of water between $300~$K and $500~$K. The blue line represents the experimental data from the NIST website \cite{NISTCWB}. The orange and green lines result from (classical) DPMD simulations trained on PBE and SCAN data, respectively. The simulations use a periodically repeated cubic box with 128 water molecules. In the simulations the box size is fixed to the experimental density \cite{NISTCWB} at each given temperature. Relative to PBE, SCAN overestimates less the experimental values, and varies less with temperature, consistent with experiment. PBE exhibits a relatively sharp conductivity maximum at around $360~$K, whereas experiment shows a broad maximum at $\approx 400~$K. The sharp PBE maximum may be an artifact of imperfect equilibration in a metastable liquid. The continuous line connects data points below the freezing temperature at $\approx 400$ K, where the PBE liquid is metastable. 
    In the Supplementary Material \cite{SupMat} the reader can find the files containing the data points for the DPMD-PBE and DPMD-SCAN simulations shown in the figure}. 
    \label{fig:kappa_T}
\end{figure}

% In the previous sections we showed that our NN approach can faithfully reproduce the \ai results, but the PBE reproduce poorly behaviour of water. 
Meta-GGA functionals like SCAN depend on the electronic kinetic energy density, in addition to the density and its gradient, making significantly 
more complicated than in the PBE case the derivation of an analytic expression for the energy flux to use in \ai MD
studies of heat transport. However, this is not necessary, as the DPMD methodology not only gives us a framework for molecular simulations having quantum-mechanical accuracy at a cost close to that of empirical force fields, but also offers us the capability of easily deriving a practical expression for the energy flux, in situations where it would be difficult to obtain it directly from first principles. To follow this route, we trained a DP model using the SCAN-DFT dataset of Ref. \onlinecite{SCANdata}. The thermal conductivity predicted by this model, at $T\approx 430~\mathrm{K}$ and at the same density used in our previous PBE simulations, is $\kappa = 0.88 \pm 0.05 $W/(mK), which is closer to experiment, but still not in perfect agreement with it. Recent studies \cite{DeepWater2021,Piaggi2021} found that the melting temperature of SCAN-DP ice Ih
models is around $310~$K, a value very close to the corresponding DFT temperature, according to perturbative estimates \cite{Piaggi2021}.      
While still not perfect, this result is far superior to PBE, whose estimated ice Ih melting temperature 
should be around $400~$K or higher \cite{Sit2005,Soohaeng2009}. Thus, one might argue that the $100~$K temperature offset used in our PBE-DFT simulations would be inappropriate here, but the rather broad temperature range displayed in \cref{fig:kappa_T} shows that the thermal conductivity of water is rather insensitive to temperature at near ambient pressure.
 
%, we can try to use a neural network potential trained on snapshot obtained via the SCAN potential, using the data set from \cite{SCANdata}.
% First of all we can compute $\kappa$ at conditions similar to those in \cref{tab:resultkappaPBE}, liquid water at the same density and $T=432~$K: $\kappa = 0.88 \pm 0.05 $W/(mK). The SCAN potential gives a result closer to the experimental value, $0.606$ W/(mK) \cite{Ramires1995}, proving again simulate better the behavior of water than PBE.  
% We performed the simulations over the range of temperatures $300$K-$500$K, for every 
The simulations reported in \cref{fig:kappa_T} have been performed by fixing the size of the simulation-box in order to match the experimental density \cite{NISTCWB} at each reported temperature. At each temperature, we first performed an NVT simulation lasting for a few dozen ps, in which the system was coupled to a Nos\'e-Hoover thermostat, followed by a $880~$ps long NVE simulation, in order to compute the thermal transport coefficient. The solid line in \cref{fig:kappa_T} connects PBE data at temperatures below $400~$K, i.e., below the estimated freezing temperature  of this model \cite{Sit2005,Soohaeng2009}. At these temperatures PBE water is sluggish and difficult to equilibrate. 
% \cref{fig:kappa_T} compares the temperature behavior of $\kappa$ for the experimental data, and the results from MD simulation via the NN potential trained over the PBE and SCAN \ai data, respectively.

%\pagebreak
SCAN overestimates $\kappa$ less than PBE, consistent with the better representation of the covalent bond length of the water molecule in the liquid provided by this functional~\cite{Chen10846}. The experimental data show a broad maximum around $400$~K, while PBE exhibits a sharp maximum around $360~$K, i.e., below the estimated freezing point of this model. The SCAN results are closer to experiment and are consistent with a broad maximum of the thermal conductivity in the explored region. 
Whether the residual discrepancy between DFT-SCAN simulations and experiment is due to a residual inaccuracy of the XC functional or to neglect of quantum effects on the nuclear motion is an issue that would require further work to be clarified.

\section{Conclusions}\label{sec:Conclusions}
In this work we have shown that DNN potentials generated according to the DP framework and properly trained on DFT data are a powerful tool to study
the transport properties of water, and likely of other material systems, with quantum-mechanical accuracy at a nearly empirical force field cost. An important byproduct of this technology is that it allows one to derive numerically practical expressions for the energy current, 
even in cases where analytical derivations from the DFT functional would be hard, as we have shown in the case of the SCAN functional.
Our results show that PBE-DFT overestimates the thermal conductivity by $\approx 60\%$. The SCAN meta-GGA functional reduces this error by approximately a factor of two, which is not quite negligible. Whether this residual discrepancy should be ascribed mostly to residual inaccuracies of the XC energy functional or to neglect of nuclear quantum effects in the particle dynamics, is an issue that deserves further study. 
As a final remark, we would like to stress that the   
method presented here should be useful in fields, such as, e.g., the geosciences and the planetary sciences, where the transport properties of different phases of matter at extreme pressure and temperature conditions, that are difficult to reproduce in the laboratory, are a key ingredient in quantitative evolutionary models of the earth and/or other planets. The reliability of such models stands in fact on the accuracy of the relevant conductivities under the thermodynamic conditions of interest \cite{Stixrude2020,Grasselli2020}.  
%In this work we presented a computation of heat transport coefficient by neural network potential. We benchmark the results against an \textit{ab initio} simulation at PBE gga level of theory, showing that the neural network can faithfully reproduce the DFT results, but drastically reducing the computation cost. The PBE proved to be not accurate enough to reproduce the complex dynamics of water, thus we applied our scheme using a new neural network potential trained over accurate SCAN data. This last potential proved to obtain thermal coefficient closer to the experimental ones but still overshoot the results. This will be investigate in futures works in order to understand if the shift from the experiments comes from the inaccuracy of SCAN potential of the missing nuclear quantum effects.

\section*{Data and code availability}\label{sec:availability}
In the Supplementary Material \cite{SupMat} the reader can find two files, \texttt{kappa\_T\_DPMD-PBE.dat} and \texttt{kappa\_T\_DPMD-SCAN.dat}, containing the data points shown in \cref{fig:kappa_T} for the DPMD-PBE and DPMD-SCAN simulations, respectively.

In the latest versions of DeePMD-kit the authors released a code to compute the heat current with the method presented in this paper. This code extends the LAMMPS \cite{PLIMPTON19951,Thompson2021,Lammpsweb} interface of DeePMD-kit allowing the computation of the heat current via the command \texttt{compute heat/flux}. For more info see the documentation on DeePMD-kit \cite{deepdoc}.

\begin{acknowledgments}
    DT, RB, and SB are grateful to Federico Grasselli for enlightening discussions throughout the completion of this work. This work was partially funded by the EU through the \textsc{MaX} Centre of Excellence for supercomputing applications (Project No. 824143). LZ and RC acknowledge support from the Center Chemistry in Solution and at Interfaces funded by the DOE Award No. DE-SC0019394. HW is supported by the National Science Foundation of China under Grant No. 11871110.
\end{acknowledgments}

\appendix

\section{Derivatives} \label{sec:derivatives}
The derivative of the local energy, $\frac{\partial w_m}{\partial \bm r_n}$, is a key component in the computation of the energy flux, \cref{eq:J^e}.
%\DTnote{to answer the 4 point of second referee} \DTcancel{As already mentioned in }\cref{sec:deepMD}\DTcancel{, it is composed of two terms, i.e.,
%$\frac{\partial w_m}{\partial \tilde {R}_m}$ that can be computed directly by tensorflow \cite{tensorflow2015-whitepaper}, and $\frac{\partial \tilde {R}_m}{\partial \bm {r}_n}$, which must be computed explicitly }\cite{NIPS2018_7696,Linfeng2018}.
As already mentioned in \cref{sec:deepMD}, it is composed of two terms, i.e.,
$\frac{\partial w_m}{\partial \tilde {R}_m}$ and $\frac{\partial \tilde {R}_m}{\partial \bm {r}_n}$. Since $w_n$ is a well defined and easy to compute function of the \emph{local environment matrices} $\tilde{R}_m$ \cite{NIPS2018_7696}, the first term can be easily obtained from TensorFlow \cite{tensorflow2015-whitepaper} using the same back-propagation approach that is commonly used during the training of a DNN \cite{Goodfellow-et-al-2016,Rumelhart1986LearningRB}. The second term must, instead, be computed explicitly \cite{NIPS2018_7696,Linfeng2018}.
Given the definition in \cref{eq:Rtilde} and the following smoothing function:

 \begin{align}\label{eq:smoothing}
\sigma(r_{mn})&=	\begin{cases}
 1 \quad r_{mn}<r_{c1} \\
 -6\Omega^5 + 15\Omega^4-10\Omega^3+1  \quad r_{c1}<r_{mn}<r_c \\
0 \quad r_{c}<r_{mn}
\end{cases}
\end{align}
where $r_{c1}$ is the smoothing cut-off radius and $\Omega=\frac{r_{mn}-r_{c1}}{r_c-r_{c1}}$, 
we get by applying the chain rule:
%Introducing the notation $r_{q(i)m}$ to denote distance between the atoms $m$ and its neighbour corresponding to the $i$-th line of matrix $\tilde{R}_{m}$ and applying the chain rule we get:
\begin{equation}
\frac{\partial \tilde{R}_m}{ \partial r_{n}^{\tau} } = \frac{\partial \tilde{R}_m}{ \partial r_{ql}^{\gamma} } \frac{\partial r_{ql}^{\gamma}}{\partial r_n^{\tau}}
\end{equation}
where sums on repeated indices are implied, and $\tau,\gamma = 1,2,3 \equiv x,y,z $ denote Cartesian coordinates. We find: 

\begin{align}
 \frac{\partial r_{ql}^{\gamma}}{\partial r_n^{\tau}}&= \delta_{\gamma,\tau}(\delta_{n,q}-\delta_{n,l}) \\
  \frac{\partial \tilde{R}_m}{ \partial r_{ql}^{\gamma} } & = \frac{\partial \tilde{R}_m}{ \partial r_{qm}^{\gamma}}\delta_{l,m} + \frac{\partial \tilde{R}_m}{ \partial r_{ml}^{\gamma}}\delta_{q,m} 
\end{align}
where $\delta_{nm}$ is the Kronecker delta. 

Using $i$, $j$ to represent line and column indices of the element of $\tilde{R}_m$ to be differentiated, a general element of $\left[ \frac{\partial \tilde{R}_m}{\partial r_{qm}^{\gamma}} \right]_{ij}$ is non-zero only if atom $q$ is the $i-th$ neighbour of $m$ in the matrix $\tilde{R}_m$:

\begin{equation}\label{ElemdR_dr}
\left[ \frac{\partial \tilde{R}_m}{\partial r_{qm}^{\gamma}} \right]_{i,j}=\begin{cases}
 \frac{ r_{qm}^{\gamma}}{r_{qm}^2}\left ( \frac{\partial \sigma_{qm}}{\partial r_{qm} }- \frac{\sigma_{qm}}{r_{qm}} \right )  \quad \text{if } j=1 \\
\begin{aligned}
& \frac{\partial \sigma_{qm}}{\partial r_{qm}} \frac{r_{qm}^{\gamma} r_{qm}^{j-1}}{r_{qm}^3}- 2\sigma\frac{r_{qm}^{\gamma} r_{qm}^{j-1}}{r_{qm}^4} \\
&+  \delta_{\gamma,j-1} \frac{\sigma_{qm}}{r_{qm}^2}  \quad \text{if } j\neq 1
\end{aligned}
\end{cases}
\end{equation}
where $\sigma_{nm} =\sigma(r_{nm})$. With the same approach a similar expression for $\left[ \frac{\partial \tilde{R}_m}{\partial r_{ml}^{\gamma}} \right]_{i,j}$ can be obtained.

\section{Neural network training}\label{sec:NeuralNetworkTraining}

\subsection{\label{sec:Traing}Training parameters}
%In order to obtain the training set of data we performed AIMD simulations of 125 water molecules at $\approx 406 $K, $423$K, $500$K, $1000$K contained in a cubic periodic cell of side $15.51$ \AA$~$, with a fictitious electronic mass of 25 atomic units (a.u.) and a time step $\Delta t = 0.2 $fs. The simulations are performed via the \verb|CP| package of \qe\ \cite{Giannozzi_2009,Giannozzi_2017} at the %PBE0-TS level
%PBE level, with the HSCV norm-conserving pseudopotentials \cite{Hamann2013}.
The NN PBE model in \cref{Water_properties} is constructed with the DeePMD-kit \cite{Wang2017} and the present appendix contains the main parameters of the model.
In the definition of the \textit{local environment matrices}, the two radii inside the smoothing function in \cref{eq:smoothing} are $r_{c1}=3.50~$\AA and $r_c=7.00~$\AA. The embedding network has three layers with 25, 50 and 100 neurons respectively, whereas the fitting network has three layers with 240 neurons each. %We adopt a ResNet-like architecture \cite{He_2016_CVPR} in both cases and 
The loss function is optimized using the Adam stochastic gradient descent method \cite{Kingma2014}, with a learning rate starting at $0.005$ and exponentially decaying, with a decay rate of $0.98$, every $10^5$ training step for a total of $1.5\cdot 10^6$ training steps. In order to optimize training the coefficients $p_E$ and $p_f$ in \cref{eq:Loss} were adjusted, respectively, from 0.05 to 1, and from 1000 to 1, during training.
%After the training we performed a classical MD simulation $300~$ps long (with timestep $dt=0.2~$fs) with 125 H$_2$O molecules at $430~$K and $1.0~$atm.

%\section{\label{NNtest}Testing}
%Once we trained the neural network potential we checked the accuracy of the model testing the results on a batch of validation data of $N_v = 800$ snapshots of 125 molecules of water at temperature between $400~$K and $1000~$K.
%We compared the \ai forces with the ones predicted by the NN potential, and the average absolute error $\bar{\epsilon} = \frac{1}{3N_vN}\sum_{b=0}^{N_v}\sum_{i=0}^{N}\sum_{\alpha=1}^3 \vert F^{\text{DFT}}_{b,i,\alpha} - F^{\text{NN}}_{b,i,\alpha}\vert = 0.051~$eV/\AA (for more details see \cref{sec:validation}).

\subsection{Training test}\label{sec:validation}
\begin{figure}[tbh]
  \centering
  \includegraphics[width=\linewidth]{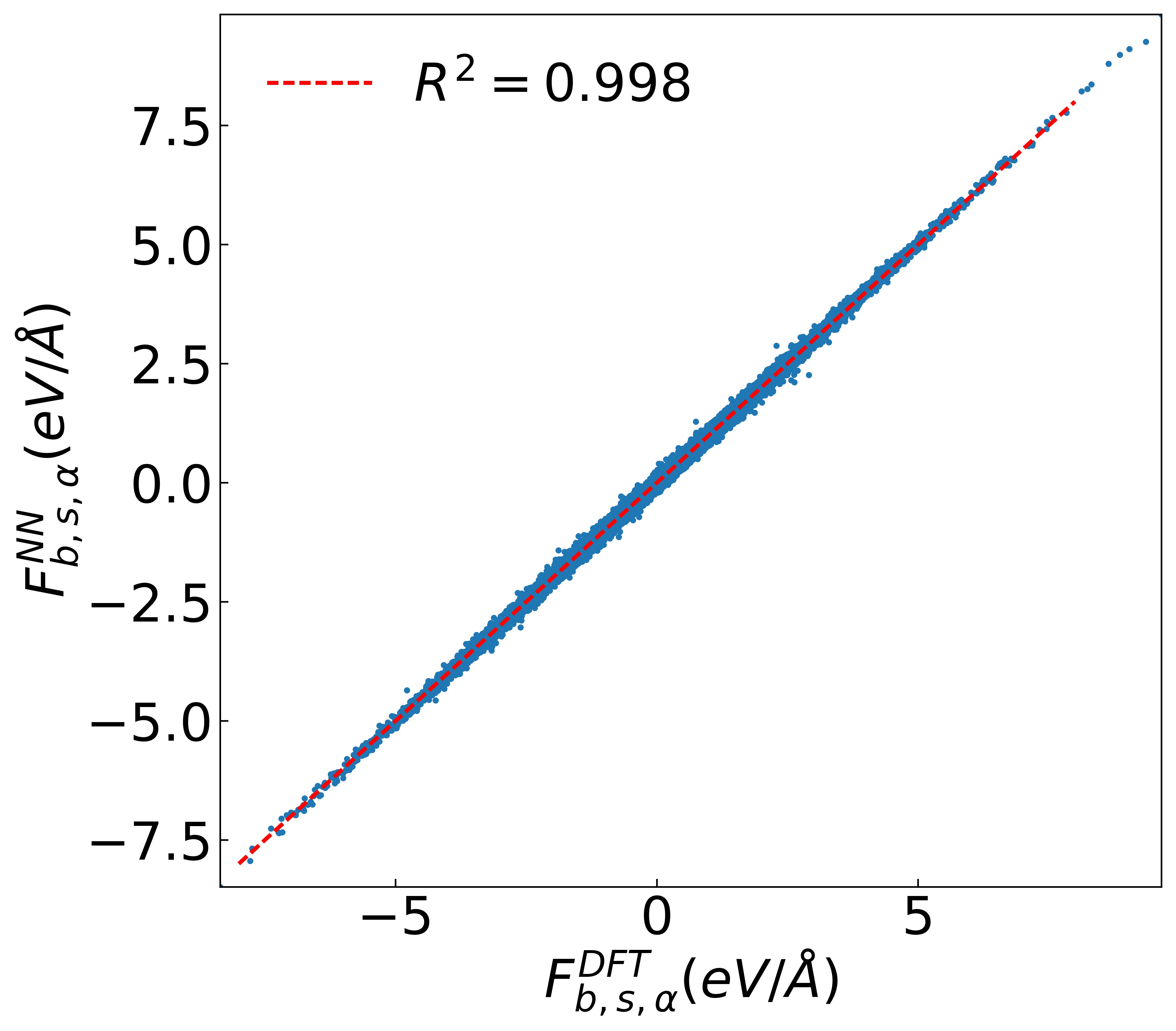}
  \caption{Direct comparison between the \ai force components and the corresponding NN prediction. The indexes $b$, $s$, $\alpha$ (see main text) label, respectively, the snapshot, the atom, and the Cartesian coordinate of the force. The red dashed line represent $F^{\text{DFT}}_{b,s,\alpha} = F^{\text{NN}}_{b,s,\alpha}$, that fits the data with $R^2=0.998$.
  }
  \label{fig:forcesvs}
\end{figure}

The PBE neural network was tested against a set of $N_v = 800$ independent snapshots of 125 molecules of water at temperatures in the range [400 K -- 1000K], obtaining a root-mean-square error of the forces of $0.05~$eV/\AA. 
\cref{fig:forcesvs} shows a direct comparison
between the $\alpha$ component of the \ai force for the
$s$-th atoms in the $b$-th snapshot and the corresponding NN prediction. The red dashed line correspond to $F^{\text{NN}}_{b,s,\alpha} = F^{\text{DFT}}_{b,s,\alpha}$, that fits the data with a \textit{coefficient of determination} $R^2=0.998$. $R^2$ is computed with the usual formula for linear regression: 
\begin{equation}
    R^2=1-\frac{\sum_i (F^{\text{DFT}}_i-F^{NN}_i)^2}{\sum_i (F^{\text{DFT}}_i-\bar{F}^{DFT})^2},
\end{equation} where $\bar{F}^{DFT}$ is the average of all the force components in the dataset. 

%\begin{figure}[htb]
%    \centering
%    \includegraphics[width=\linewidth]{forceerr_histo.pdf}
%    \caption{Histogram of the absolute error of the force %components, $F^{\text{DFT}}_{i,\alpha} - %F^{\text{NN}}_{i,\alpha}$ where index $i$ select the $i$-th %atom and $\alpha$ select the Cartesian coordinate x, y, z. The %average absolute error $\bar{\epsilon} = 0.03999~$eV/\AA  }
%    \label{fig:forceshisto}
%\end{figure}

\subsection{\label{Benchmark}Benchmark of water properties}
\begin{figure}[htb]
    \centering
    \includegraphics[width=\linewidth]{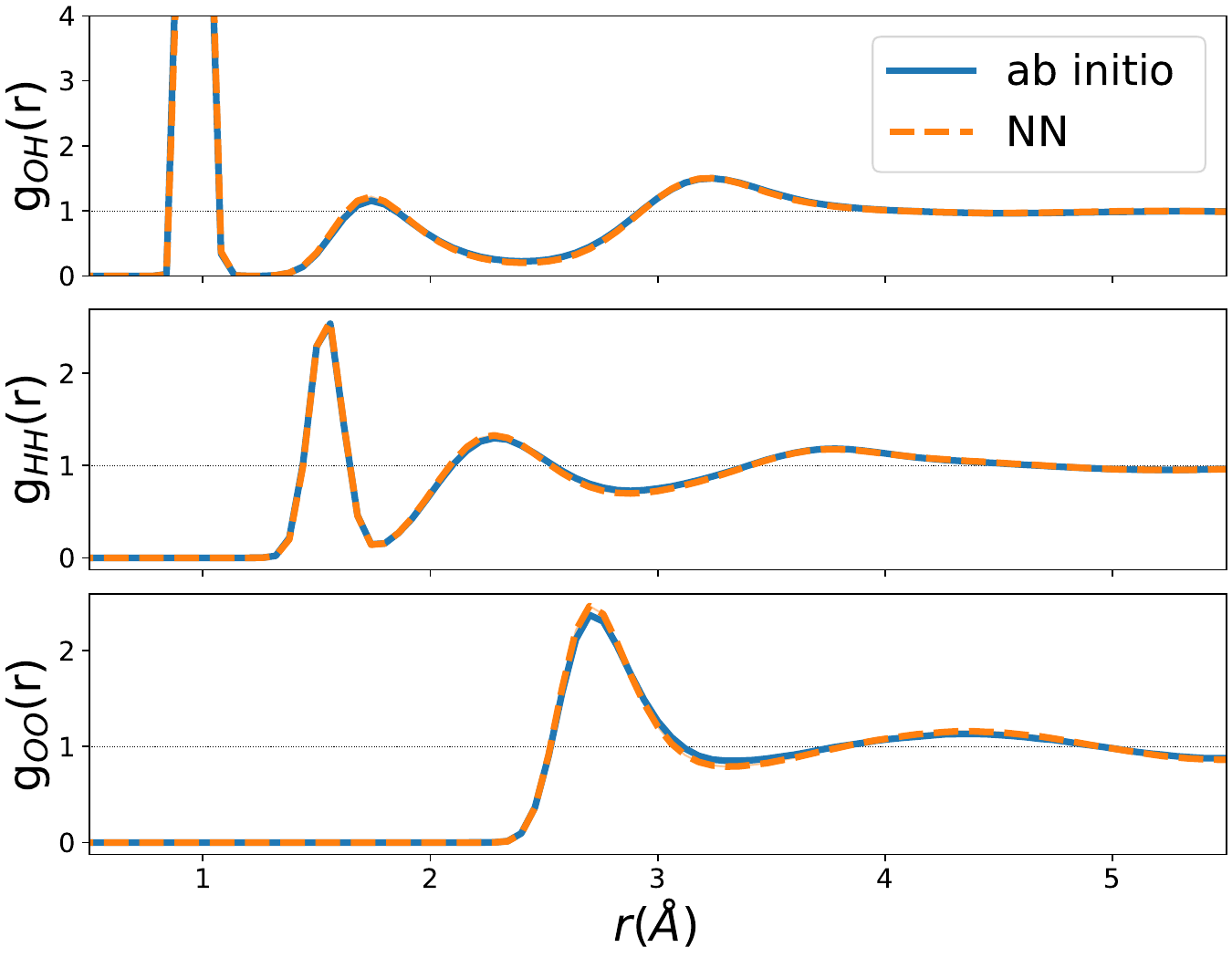}
    \caption{Comparison of the radial distribution functions of liquid water from \ai (continuous blue line) and DP (dashed orange line) simulations,
    respectively. More details on the simulations can be found in the main text.}
    \label{fig:RDF_liquid}
\end{figure}

\begin{figure}[htb]
    \centering
    \includegraphics[width=\linewidth]{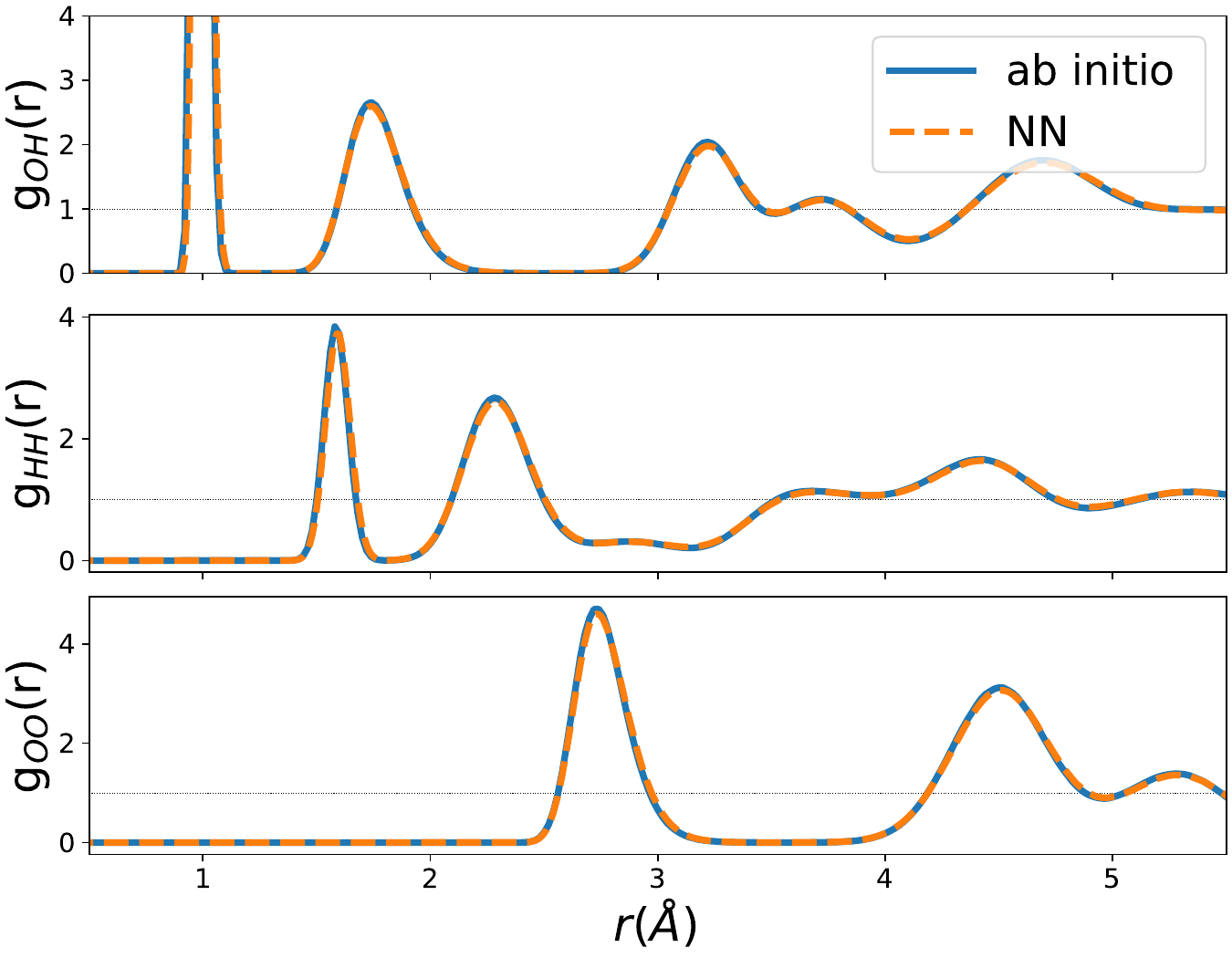}
    \caption{Comparison of the radial distribution functions of ice Ih from \ai (continuous blue line) and DP (dashed orange line) simulations,
    respectively. More details on the simulations can be found in the main text.}
    \label{fig:RDF_ice}
\end{figure}

To estimate the quality of the trained DP model we compared some simple static and dynamical properties of the model with their \ai counterparts. We ran DPMD simulations of water at the same thermodynamic conditions of the \ai simulations reported in \cref{sec:abinitio_results}.
\cref{fig:RDF_liquid,fig:RDF_ice} compares the oxygen radial distribution
functions, $g(r)$, from DP and \ai simulations
of liquid water (third and seventh line of \cref{tab:resultkappaPBE}), and of ice-Ih (fourth and last line of \cref{tab:resultkappaPBE}).
Both structures are well described by the DP model. This is true also for the ice-structure even though no
ice-snapshots were included in the training data set.

\begin{figure}[htb]
    \centering
    \includegraphics[width=\linewidth]{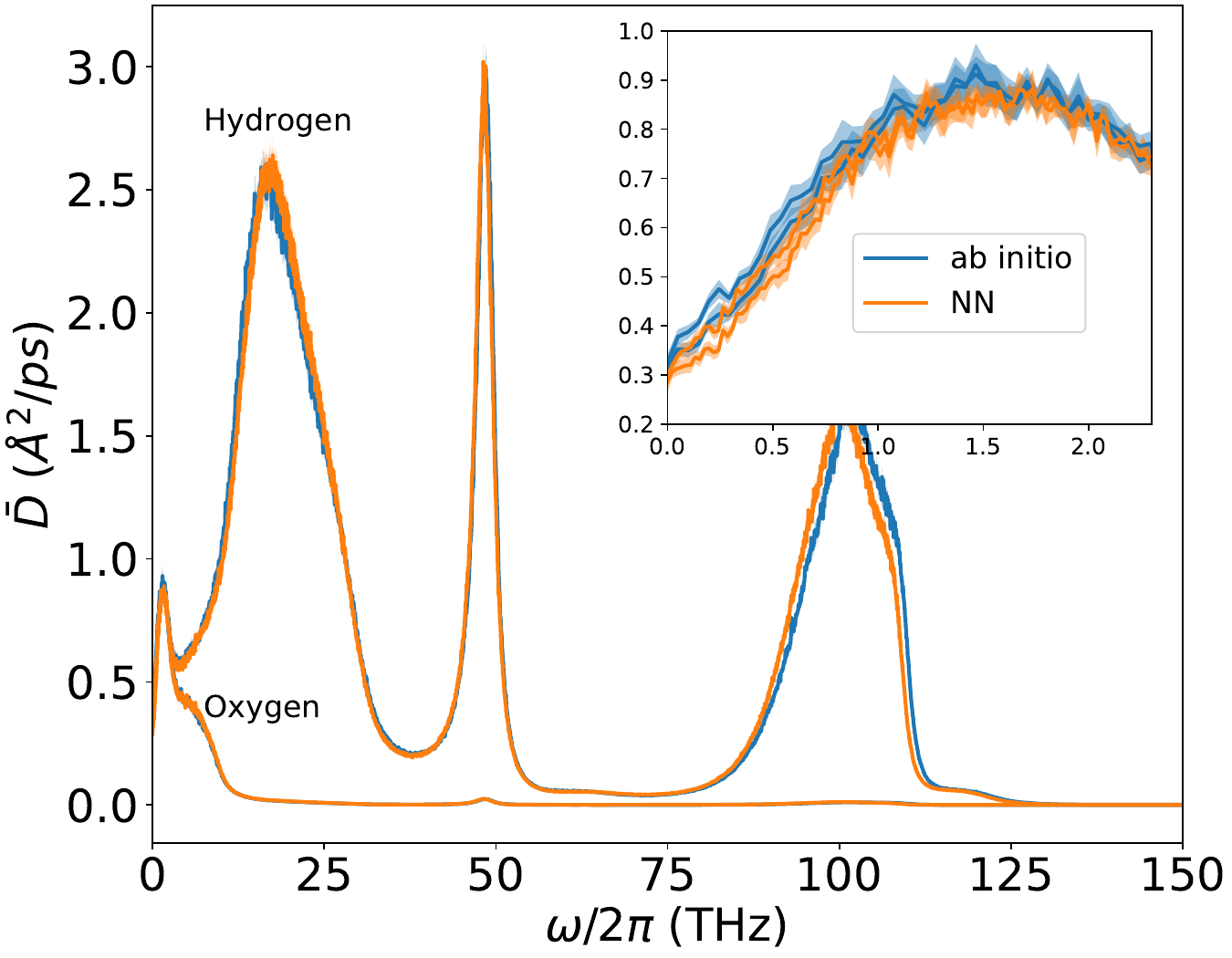}
    \caption{Comparison of the oxygen and hydrogen velocity power spectra of liquid water from \ai (blue line) and DP (orange line) simulations, 
    respectively. The simulations used the same periodic cubic cell with density $\rho=1.00~$g/cm$^3$ containing $125$ water molecules, at $\approx 410K$. The inset shows the region near $\omega=0$ used to estimate the diffusivity.}
    \label{fig:spectrum}
\end{figure}

For liquid water, we computed also the power spectra of the oxygen and hydrogen velocities \cref{eq:D_omega}, respectively, and their zero frequency values, the diffusion coefficients.
\cref{fig:spectrum} shows the power spectra of liquid water systems mentioned above. It can be seen that DP and \ai models give consistent diffusivities (see \cref{tab:resultkappaPBE} for a complete comparison of the results):
$D_{H}^{\text{AIMD}}=0.325 \pm 0.018~$\AA$^2$/ps,
$D_{H}^{\text{NN}}=0.29 \pm 0.02~$\AA$^2$/ps,
$D_{O}^{\text{AIMD}}= 0.29 \pm 0.02~$\AA$^2$/ps and
$D_{O}^{\text{NN}}=0.32 \pm 0.02~$\AA$^2$/ps.

\section{Cepstral analysis of the flux time series}
In the present work the thermal conductivity is computed via the cepstral analysis of the energy flux, as implemented in the \sportran\ code \cite{SporTran}. This technique provides a very accurate and reliable estimate of the transport coefficients and their statistical accuracy, depending only on two parameters: the effective Nyquist frequency, $f^*$, used to limit the analysis to a properly defined low-frequency window, and the number $P^*$ of cepstral coefficients. For a detailed explanation of the method and the meaning of the parameters the reader may consult \cite{Baroni2018,Grasselli2021,Ercole2017}. \cref{tab:resultkappaPBE_PeF} contains the parameters used to obtain the values of $\kappa$ in \cref{tab:resultkappaPBE}.

\begin{table}[h!]
    \centering
    \begin{tabular}{c|c|c|c|c}
     & phase   & T & $f^*$ & $P^*$  \\
     &             & K & THz & \\
        \hline 
        \multirow{4}{*}{DPMD} & liquid  & $516 $ & $9.9 $ & $11$  \\
        & liquid  & $423 $ & $17.8$ & $12$   \\
       & liquid  & $408 $ & $36.7$ & $17$   \\
         & ice Ih & $270$ & $25$ & $93$  \\
         \hline
        \multirow{4}{*}{\ai} & liquid  & $521 $ & $20.7 $ & $55$  \\
        & liquid  & $431$ & $20.1$ & $17$ \\
       & liquid  & $409$ &$45.9$ & $33$  \\
        & ice Ih  & $260$ & $30.3$ & $53$ \\
    \end{tabular}
    \caption{Table with the value of $f^*$ and $P^*$ used to obtained the values in \cref{tab:resultkappaPBE}. }
    \label{tab:resultkappaPBE_PeF}
\end{table}

%\pagebreak

\section{Size scaling for SCAN neural network potential}

\begin{figure}[hbt]
     \centering
     \includegraphics[width=\linewidth]{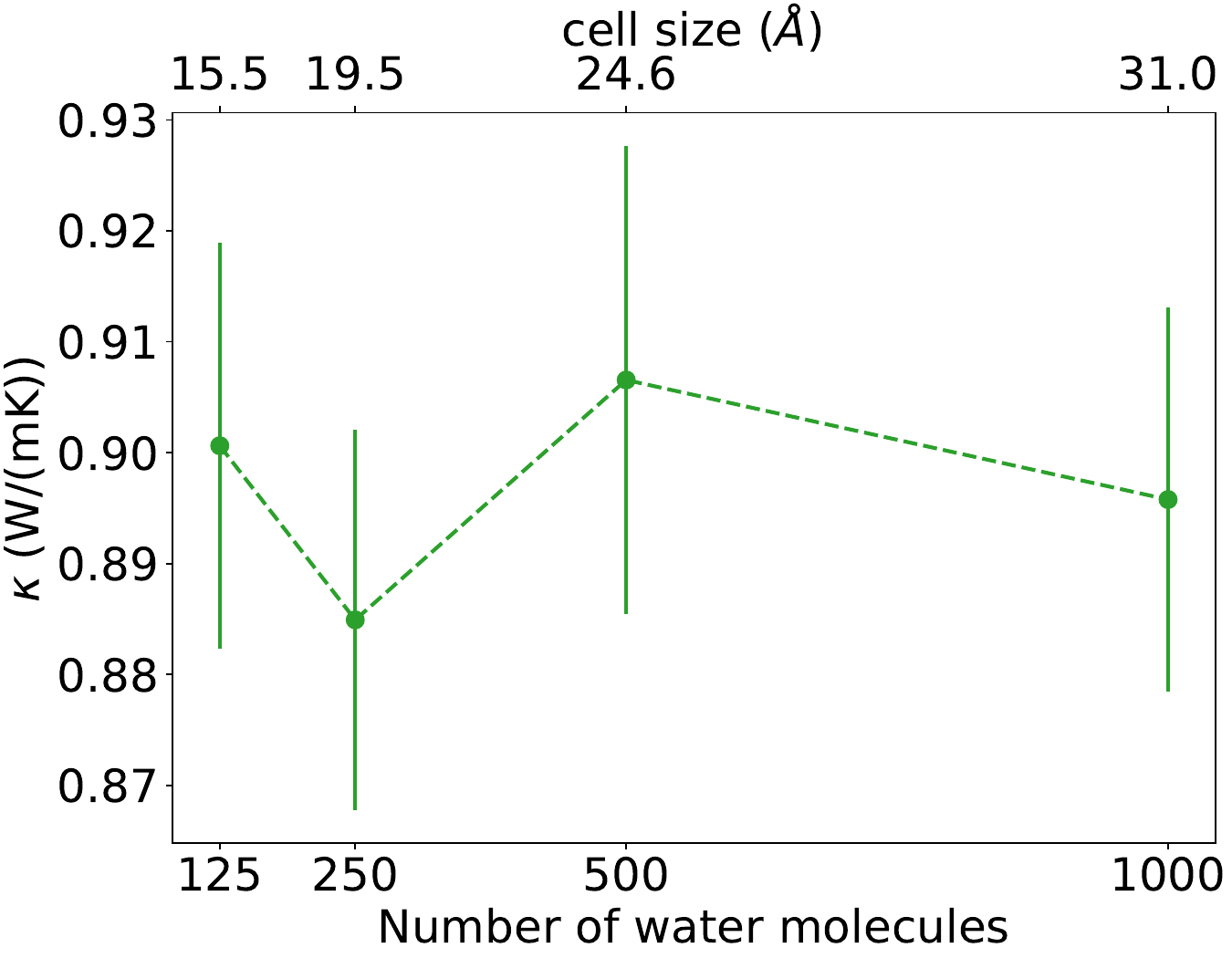}
     \caption{ The size dependence of the thermal transport coefficient $\kappa$ for simulation with the SCAN neural network potential. The test shows that no relevant size scale dependence is observed. All the quantities are evaluated from $\approx 2 $ns long trajectories. }
     \label{fig:size_scaling}
 \end{figure}

Size effects may affect the transport properties calculated in numerical simulations \cite{Yeh2004,Puligheddu2020}. In order to quantify these effects, we run $2~$ns long NVE simulations at $\approx 407~$K of SCAN-DP water at fixed density and increasingly larger cells (with up to $1000$ molecules). The results, reported in \cref{fig:size_scaling}, suggest that $\kappa$ shows no size dependence within the error bars of the simulation.   

\pagebreak
\bibliography{biblio}% Produces the bibliography via BibTeX.

\end{document}